\documentclass[journal,twoside,web]{ieeecolor}
\usepackage{generic}
\usepackage{cite}
\usepackage{amsmath,amssymb,amsfonts}
\usepackage{algorithmic}
\usepackage{graphicx}
\usepackage{textcomp}
\usepackage{multirow}
\usepackage{tabularx,booktabs}
\usepackage{array}
\usepackage{hyperref}
\DeclareMathOperator*{\argmax}{arg\,max}

\def\x{{\mathbf x}}
\def\u{{\mathbf u}}
\def\v{{\mathbf v}}
\def\y{{\mathbf y}}
\def\v{{\mathbf v}}

\def\C{{\mathbf C}}
\def\V{{\mathbf V}}

\def\D{{\mathbf D}}

\def\H{{\mathbf H}}

\def\R{{\mathbf R}}
\def\T{{\mathbf T}}
\def\U{{\mathbf U}}

\def\1{{\mathbf 1}}
\def\R{{\cal R}}

\def\BibTeX{{\rm B\kern-.05em{\sc i\kern-.025em b}\kern-.08em
    T\kern-.1667em\lower.7ex\hbox{E}\kern-.125emX}}
\begin{document}
\title{Deep Correlation Analysis for \\  Audio-EEG Decoding}
\author{Jaswanth Reddy Katthi, \IEEEmembership{Student Member, IEEE}, and Sriram Ganapathy, \IEEEmembership{Senior Member, IEEE}

\thanks{
%Manuscript received December 19, 2019; revised February 24, 2020; accepted March 5, 2020. Date of publication March 26, 2020; date of current version May 8, 2020.
This work was supported in part by the grants from Department of Atomic Energy project 34/20/12/2018-BRNS/34088}
\thanks{Jaswanth Reddy and Sriram Ganapathy are with the Learning and Extraction of Acoustic Patterns (LEAP) laboratory, Electrical Engineering, Indian Institute of Science, C. V. Raman Avenue, Bangalore, India, 560012. (e-mail: \{jaswanthk,sriramg\}@iisc.ac.in). }}

\maketitle

\begin{abstract}
The electroencephalography (EEG), which is one of the easiest modes of recording brain activations in a non-invasive manner, is often distorted due to recording artifacts which adversely impacts the stimulus-response analysis. The most prominent {\color{black} techniques} thus far attempt to improve the stimulus-response correlations using linear methods. In this paper, we propose a neural network based correlation analysis framework that significantly improves over the linear methods for auditory stimuli. A deep model is proposed for intra-subject audio-EEG analysis based on directly optimizing the correlation loss. Further, a neural network model with a shared encoder architecture is proposed for improving the inter-subject stimulus response correlations. These models attempt to suppress the EEG artifacts while preserving the components related to the stimulus. Several experiments are performed using EEG recordings from subjects listening to speech and music stimuli. In these experiments, we show that the deep models improve the Pearson correlation significantly over the linear methods (average absolute improvements of $7.4$\% in speech tasks and $29.3$\% in music tasks). We also analyze the impact of several model parameters on the stimulus-response correlation.   
\end{abstract}


\section{Introduction}
\label{sec:introduction}
\IEEEPARstart{U}{nderstanding} the human brain has been a topic of profound interest in both science and engineering fields. One of the most common methods to perform this analysis is to measure the evoked brain response for a given stimuli and establish a relation between them. 
%The common non-invasive methods to measure brain activity include electroencephalography (EEG), magnetoencephalography (MEG) and functional magnetic resonance imaging (fMRI).
%True scientific methods came into picture very recently with the scientific revolution. One of the interesting functions of the brain is processing the auditory information collected through the binaural ears. The pipeline of the auditory processing, especially by humans, is being studied extensively. Processes involving collection of the auditory information and transceiving it to the auditory cortex are studied to an extent. But, it is still not so clear about how our brain processes this information. Studying it involves two things. One, collecting the brain data while the auditory information is being processed and the other is to make sense of the data collected. 
% The pipeline of the auditory processing, especially by humans, is being extensively studied. Though other elements, in this pipeline, that involve collecting the auditory information and transceiving the information to the auditory cortex are understood to an extent, understanding how our brain processes this information is still under the microscope. There are two things in play here: one, collecting the brain data while processing the auditory information and the other is to make sense of the data collected. 
The electroencephalography (EEG)  constitutes the simplest  non-invasive technique to collect brain signals while having sufficient temporal resolution for  auditory analysis. Since the EEG recordings involve scalp level measurements, the recordings are significantly impacted by noise~\cite{sanei2007eeg}. The most popular method for analyzing auditory invoked EEG signals is the classical event-related potential (ERP) approach~\cite{galambos1962electroencephalograph,soman2019eeg}. This approach involves averaging the  EEG responses 
%to similar stimuli conditions 
in time/frequency domain to suppress the noise in the recordings~\cite{luck2014introduction}.  However, this approach is limited to isolated stimuli that have to be repeated and is therefore often restrictive for use in the analysis of natural stimuli like speech and music. 
%Thus, new approaches for single-trial EEG data analysis are in demand for longer and naturalistic stimuli, like speech and music. 

% 

% Since they collect data from the scalp, the brain response to the presented stimuli cannot be collected exclusively. A lot of underlying processes of the brain also get recorded while collecting the desired response to our provided stimulus. As the signals collected by these techniques represent all the processes running inside the brain, processes corresponding to the stimuli cannot be isolated easily. This is one of the problems in understanding how the brain operates on a stimulus. This problem can be alleviated by repeating the experiment for a finite number of times and averaging the brain responses collected. This way, all the brain signals not related to the stimulus can be identified and subtracted from the data. This results in brain response mainly related to the stimulus only. But it is efficient to do only for short-time stimuli. It is difficult to perform for a longer and naturalistic stimulus like speech.  It would be tiresome to repeat for such stimuli. One way to tackle this problem is by aggregating brain response from different subjects. Ideally, the process should be efficient to remove the subject-specific signals and retain only stimulus related signals from the EEG.

One of the first successful attempts in this direction is the temporal response function (TRF) proposed by Lalor et al.~\cite{lalor2010neural}. The linear TRF model describes the relationship between a stimulus and its response as a linear time-invariant (LTI) system. It can be a linear forward model where the model estimates the EEG response for the stimulus or a backward model where the model predicts the components of the stimulus from the  EEG response. The model estimation is performed using linear least squares. The performance of these models is typically validated using the Pearson correlation value between the target signal and the predicted signal~\cite{crosse2016multivariate}.  

The initial studies used the slowly varying speech envelopes of the stimuli and the corresponding single-trial EEG responses~\cite{lalor2009resolving,ding2013adaptive}. The analysis can also be extended to speech spectrograms~\cite{ding2012emergence}, phonemes~\cite{di2015low}, or semantic features~\cite{broderick2018electrophysiological}. 
% In this paper, we use the low frequency speech envelopes.  

The Canonical Correlation Analysis (CCA) is an extension of the linear methods for analysis. Here,  two signals are projected onto a subspace that maximizes the correlation between them~\cite{thompson1984canonical}. It determines a set of orthogonal directions on which the two signals are highly correlated. The CCA has been shown to be better than forward and backward TRF models in auditory-EEG analysis recently~\cite{de2018decoding,alickovic2019tutorial}.

{\color{black} For each subject,  the stimulus and response representations are defined as ``views" of the auditory signal. The stimulus-view represents the audio signal using a temporal envelope and the response-view represents it using the brain responses collected as EEG recordings.} The linear CCA can be performed only on two views of the data at a time. In order to aggregate the EEG responses from multiple-views (subjects), multiway CCA (MCCA) or generalized CCA~\cite{correa2010multi,fu2017scalable,zhang2017inter} has been proposed.   As all views (EEG responses) represent the same object (audio stimulus), some components are common across the views~\cite{parra2018multi}. The application of multiway CCA for EEG mapping has shown improvements over the intra-subject CCA ~\cite{de2019multiway}. 

In this paper, we explore a deep neural network based architecture for correlating the EEG response with the stimulus features. The deep CCA framework, introduced by Andrew et al.~\cite{andrew2013deep}, had shown promise over the linear CCA for image data. However, the direct application of the deep CCA to the EEG data is cumbersome as the EEG data is significantly noisy   with signal-to-noise ratio (SNR) below $-20$dB~\cite{sanei2007eeg}. The dropout strategy~\cite{srivastava2014dropout} alleviates the impact of noise partly.
% The impact of noise can be alleviated partly with the use of the  dropout  strategy  which helps to  avoid over-fitting \cite{srivastava2014dropout}. 
%A leaky-ReLU based non-linear activation is also shown to be more robust. 
In  audio-EEG  experiments, we  show  that  the  deep  CCA consistently improves over the linear CCA model.
% (statistically significant - $p < 0.05$ for $5$ out of $6$ subjects from the speech-EEG data). 
%In this paper, we extend our previous work \cite{embc on more speech/audio datasets and with additional analysis 

We also propose an approach for deep multiway CCA where multiple EEG responses for the same stimuli can be combined in a neural network architecture. For this task, we use a reconstruction approach with a shared hidden representation to derive the deep transform that aligns multiple views. Using this novel approach, we show that the deep MCCA improves over the linear MCCA model ~\cite{de2019multiway}. 
% It is also found that the improvements are statistically significant on all subjects ($p<0.05$). 
In subsequent analysis, we also illustrate how the deep MCCA can be combined with the deep CCA model for EEG analysis.

The data used in the  experiments consists of EEG responses for speech and music listening tasks. The speech task is the same dataset used in the linear CCA work by Cheveigné et al.  \cite{de2018decoding} where subjects listen to the narration of an audiobook. The music dataset used in this work is the Naturalistic Music EEG Dataset - Hindi (NMED-H) \cite{kaneshiro2016naturalistic} which is an open dataset of EEG  responses collected for Hindi pop songs.

% It is an extension to the linear CCA. It extracts the components that are common among all the signals provided to the model. Let, there are multiple representations (views or dataviews) of a particular object, which can be obtained from different point of views.  And those are the signals MCCA tries to extract. But, MCCA also works on linear assumptions of the system. With its inspiration, we have extended our Deep CCA framework to extract common features from multiple subjects.
 
%In this work, we discuss about how linear and deep versions of CCA and MCCA operate on multiple dataviews and their performance on  audio-EEG data.
 
 The remainder of the paper is arranged as follows. Section~\ref{sec:related_work} highlights prior work done in the domain.  Section~\ref{sec:background} describes the background of  linear  CCA and multiway CCA. Section~\ref{sec:deepCCA} discusses the proposed deep CCA model and the deep MCCA model. Section~\ref{sec:Setup} details the datasets and experimental setup. Section~\ref{sec:results} reports the results of the proposed deep models and a comparison with linear models. Section~\ref{sec:discussion} presents a discussion on the hyper-parameters. Finally, Section~\ref{sec:summary} presents the summary of this work.
% and how dropout regularization helps us to overcome overfitting of the deep models to the noise present in EEG.

\section{Related Prior Work}\label{sec:related_work} 

Machine learning methods for the extraction of information from brain signals like EEG have a significant impact on both understanding and applications like brain-computer interfaces (BCI)~\cite{schirrmeister2017deep}.  It is therefore of profound importance to transfer the recent advancements in machine learning (for example, deep learning~\cite{lecun2015deep}) to improve the models for brain signal decoding and single-trial analysis. One of the first works in this direction involved the use of convolutional neural networks to identify the P300 wave in EEG signals~\cite{cecotti2010convolutional}. The recent years have seen the use of deep learning for several brain mapping tasks like computational memory prediction~\cite{sun2016remembered}, driver's cognitive state prediction~\cite{hajinoroozi2016eeg}, and the brain activity reconstruction for visual stimuli~\cite{yuan2019wave2vec}. A review of several efforts in decoding brain activity using deep learning techniques is given in Zheng et al.~\cite{zheng2020decoding}. 

In auditory tasks, EEG recordings have shown to contain rhythm information in music perception using classifiers based on deep networks~\cite{stober2014using}. A recent work by Das et al.~\cite{das2020linear} has shown that auditory attention decoding in the perception of noisy speech can also be improved by deep learning techniques.  In multi-speaker cocktail party scenarios, Deckers et al.~\cite{deckers2018eeg} showed that neural networks are capable of identifying the attended speaker.  
A Convolutional Neural Network (CNN) based model for EEG-based speech
stimulus reconstruction was also proposed  by de Taillez et al.~\cite{de2020machine}, showing that deep learning is a feasible
alternative to linear decoding methods. Our work on deep CCA models~\cite{katthi2020deep} for intra-subject analysis and inter-subject analysis~\cite{katthi2021deep} is extended in this paper with inter-subject models and additional evaluations on music dataset.

%In this paper, we propose a neural framework for developing models that correlate the EEG recordings with the audio stimuli. We propose two models - the deep canonical correlation analysis (DCCA) and deep multiway canonical correlation analysis (DMCCA). The DCCA model attempts to transform audio and EEG signals from a single subject while the DMCCA model is a preprocessing step which aligns the EEG recordings from multiple subjects for the same stimulus. %We show that the proposed models advance linear methods proposed recently like the CCA~\cite{de2018decoding} and multiway-CCA~\cite{de2019multiway} and provide statistically significant improvements over the linear counterparts for  both speech and audio tasks. 

\section{Mathematical Background}\label{sec:background} 

\subsection{Linear Canonical Correlation Analysis}
For a dataset with pairs of multi-variates, linear Canonical Correlation Analysis~\cite{hotelling1992relations} obtains an optimal linear transform for both the views such that the Pearson correlation of the two transformed vectors is maximized. Let $ \mathbf{x} \in \mathcal{R}^{\mathcal{D}_1}$ and $\mathbf{y} \in \mathcal{R}^{\mathcal{D}_2}$ {\color{black} be two random vectors that  represent two views of the data}. Let $d$ be the dimension of the projected subspace. The subspace is determined such that resultant projection vectors are maximally correlated. For example, if $d=1$, then a pair of transform vectors $\mathbf{u}_1 \in \mathcal{R}^{\mathcal{D}_1}$ and $\mathbf{v}_1 \in \mathcal{R}^{\mathcal{D}_2}$ need to be determined such that $ {\hat{x}} = \mathbf{u}_1^\top\mathbf{x}$ and ${\hat{y}} = \mathbf{v}_1^\top\mathbf{y}$ are maximally  correlated. Mathematically,  
\begin{align}
    \color{black}{\rho}\left(\mathbf{u}_1, \mathbf{v}_1\right) &= \frac{\mathbf{u}_1^\top\mathbf{C_{xy}}\mathbf{v}_1}{\sqrt{\mathbf{u}_1^\top\mathbf{C_{xx}}\mathbf{u}_1\mathbf{v}_1^\top\mathbf{C_{yy}}\mathbf{v}_1}} \\ 
     (\mathbf{u}_1^*, \mathbf{v}_1^*) &=  \argmax_{\mathbf{u}_1, \mathbf{v}_1} \color{black}{\rho}\left(\mathbf{u}_1, \mathbf{v}_1\right) \nonumber 
\end{align}

where, $\mathbf{C_{xx}}$, $\mathbf{C_{yy}}$ are the auto-correlation matrices of $\mathbf{x}$, $\mathbf{y}$ respectively while  $\mathbf{C_{xy}} = \mathbf{E}[(\mathbf{x} - \mathbf{\mu_x}) (\mathbf{y} - \mathbf{\mu _y})^\top]$ is the cross correlation matrix. Here,  $\mathbf{\mu_x}$ and $\mathbf{\mu_y}$ are the mean vectors of $\mathbf{x}$ and $\mathbf{y}$ respectively.  

It can be shown that the optimal solution to the transform vectors $\mathbf{u}_1^*$ and $\mathbf{v}_1^*$ is given by the first left and right singular vectors of the matrix 
\begin{align}
    \mathbf{T} \triangleq \mathbf{C_{xx}}^{-1/2}\mathbf{C_{xy}}\mathbf{C_{yy}}^{-1/2} 
    \label{T}
\end{align}
respectively. For $d > 1$,  the solution is obtained by finding the subsequent singular vectors of $\mathbf{T}$~\cite{andrew2013deep}. 
%Let $\mathbf{U}_k \in \mathcal{R}^{k \times \mathcal{D}_1}$ denote the linear transform for the view $\mathbf{x}$ and let $\mathbf{V}_k \in \mathcal{R}^{k \times \mathcal{D}_2}$ denote the the same for view $\mathbf{y}$ such that the the projected vectors $ \mathbf{\hat{x}}, \mathbf{\hat{y}} \in \mathcal{R}^{k}$ are maximally correlated.
% The two transform matrices $\mathbf{U}_k$ and $\mathbf{V}_k$ can be obtained by solving the following optimization problem: 
% \begin{equation}
% (\mathbf{U}_k^*, \mathbf{V}_k^*) = \argmax_{\mathbf{U}_{k}^{T} \mathbf{C}_{xx} \mathbf{U}_{k} = \mathbf{V}_{k}^{T} \mathbf{C}_{yy} \mathbf{V}_{k}= \mathbf{I}} \operatorname{Tr}\left(\mathbf{U}_{k}^{T} \mathbf{C}_{xy} \mathbf{V}_{k}\right)
% \end{equation}
%The columns of each projection matrix are also orthogonal to each other. 
%The solution ($\mathbf{U}_k^{*}$, $\mathbf{V}_k^{*}$) is top $k$ left and right singular vectors of the matrix $\mathbf{T}$.

\subsection{Linear Multiway CCA}\label{sec:mcca}
The multiway CCA is a linear method that generalizes the linear CCA to multiple (more than two) data-views. It finds a linear transform for each data-view, such that all the projections are maximally correlated to each other~\cite{parra2018multi, de2019multiway}.

Let $\mathbf{x}_n \in \mathcal{R}^{d_n}$, for $n=1\text{ to }N$, denote the $N$ data-views and $D_N = \sum_{n=1}^{N}{d_n}$. For the $1$-D projection case,  let $\mathbf{v}_n \in \mathcal{R}^{d_n}$ denote the transform vector that projects $\mathbf{x}_n$ onto the common subspace. The goal of MCCA is to find the transform vectors $\{\mathbf{v}_n\}_{n=1}^{N}$ such that the inter-set correlation (ISC)   is maximized. The ISC is defined as 
\begin{equation}\label{eq:isc}
    {\color{black}{\rho}_\text{ISC} }=\frac{1}{N-1} \frac{r_{B}}{r_{W}}
\end{equation}
where $r_B$ is the between-set covariance  and $r_W$ is the within-set covariance.  %The factor \((N-1)^{-1}\) constraints the correlation to \(\rho \leq 1\). 
The between-set and within-set covariances are,
\begin{equation}
    r_B  =  \sum_{j=1}^{N} \sum_{k=1, k \neq j}^{N} \mathbf{v}_{j}^{\top} \mathbf{R}^{j k} \mathbf{v}_{k} ~~~;~~~
    r_W  =  \sum_{j=1}^{N} \mathbf{v}_{j}^{\top} \mathbf{R}^{j j} \mathbf{v}_{j} \nonumber
\end{equation}
where $\mathbf{R}^{j k} \in \mathcal{R}^{d_j \times d_k}$ is the cross-covariance matrix between the views $\mathbf{x}_j$ and $\mathbf{x}_k$. 
% where $\mathbf{R}^{j k}=\sum_{i}\left(\mathbf{x}_{j}^{i}-\overline{\mathbf{x}}_{j}^{*}\right)\left(\mathbf{x}_{k}^{i}-\overline{\mathbf{x}}_{k}^{*}\right)^{\top}$ is the cross covariance matrix between $\mathbf{x}_j$ and $\mathbf{x}_k$. And $\overline{\mathbf{x}}_{j}^{*}$ is the mean of $\mathbf{x}_j$.

The cross-covariance matrices among all the views form elements of a block matrix $\mathbf{R} \in \R^{D_N \times D_N}$ such that $\left[\mathbf{R}\right]_{ij} = \mathbf{R}^{ij}$. By considering only the autocovariance matrices, a block-diagonal matrix $\mathbf{D} \in \R^{D_N \times D_N}$ is formed whose block diagonal entries are the same as that of $\mathbf{R}$. The optimal transform vectors $\{\mathbf{v}_n\}_{n=1}^{N}$ are obtained by solving the equation~\cite{parra2018multi}: 
\begin{equation} \label{eq:mcca_eigen}
    \mathbf{R} \mathbf{v}=\lambda \mathbf{D} \mathbf{v} 
\end{equation}
The eigenvector $\mathbf{v}\in\mathcal{R}^{D_N \times 1}$ with the maximum eigenvalue is the concatenation of the transform vectors $\{\mathbf{v}_n\}_{n=1}^{N}$. 
% matrix $\mathbf{R}$ is a block matrix with its elements as $[\mathbf{R}]_{ij} = \mathbf{R}^{ij}$  and the diagonal-block matrix $D$ whos elements are $[\mathbf{D}]_{ii} = \mathbf{R}^{ii}$.

% \begin{align}
% \mathbf{R} &=\left[\begin{array}{cccc}\mathbf{R}^{11} & \mathbf{R}^{12} & \cdots & \mathbf{R}^{1 N} \\ \mathbf{R}^{21} & \mathbf{R}^{22} & \cdots & \mathbf{R}^{2 N} \\ \vdots & \vdots & \ddots & \vdots \\ \mathbf{R}^{N 1} & \mathbf{R}^{N 2} & \cdots & \mathbf{R}^{N N}\end{array}\right], \\
% \mathbf{D} &=\left[\begin{array}{cccc}\mathbf{R}^{11} & 0 & \cdots & 0 \\ 0 & \mathbf{R}^{22} & \cdots & 0 \\ \vdots & \vdots & \ddots & \vdots \\ 0 & 0 & \cdots & \mathbf{R}^{N N}\end{array}\right]
% \end{align}

For projection onto a higher dimensional subspace $d$ ($>1$), the transform matrix for each multivariate $\mathbf{x}_n$ becomes $\mathbf{V}_n \in \mathcal{R}^{d_n \times d}$. This involves finding the top $d$ eigenvectors  of Equation~(\ref{eq:mcca_eigen}). 

\subsection {Understanding CCA for EEG Decoding}
The CCA model attempts to find a subspace of brain activity which is maximally correlated with the auditory stimulus. The EEG signal in the form of spatial channels (electrodes) and time-domain lags are used while the time-lagged audio envelopes are used as stimulus features. These two vectors form the data-views for CCA~\cite{de2018decoding}. {\color{black} Specifically, the stimuli features, $\x$, in our experiments represent the time-lagged envelope of the audio signal while the response features, $\y$, represent the  EEG recordings.
% from the scalp electrodes.  
These two features are provided at the same sampling rate and the CCA model attempts to relate them. This is done by finding the linear transforms, $\u_1^*$ and $\v_1^*$, that transform the stimulus and the response respectively, in such a way as to maximize the correlation.}

In this case, the components of CCA can also be regarded as spatio-temporal filters applied on the EEG data and modulation filters on the audio envelope. The multiple CCA components of the audio signal correspond to FIR filtered envelopes that are orthogonal. Similarly, the CCA components of the EEG signal represent spatio-spectrally filtered projections that are orthogonal.  
The CCA was used recently for auditory and audio-visual EEG analysis by Dmochowski et al.~\cite{dmochowski2018extracting}. The use of CCA in forward and backward models with time lags has shown additional improvements in correlation values~\cite{de2018decoding}.    

When multiple subjects are presented with the same stimulus, the functional similarity is expected to generate similar responses~\cite{lankinen2014intersubject}. However,  the position or orientation of the sources with respect to the electrodes may be different for different subjects and therefore, the direct mapping of the EEG responses for the same stimulus from different subjects is cumbersome. The multi-CCA attempts to align the data from each subject to a common representation that makes it possible to compare across subjects. This is achieved by deriving spatial filters that are specific to each subject~\cite{haxby2011common}.

{\color{black} For the MCCA model, the $N$ subjects' response features (EEG recordings), $\mathbf{x}_n$ for $n=1 \text{ to } N$ and the common stimulus features (audio envelope) $\x_{N+1}$, are provided as inputs. Now, the model provides a linear transform for each of them, $\mathbf{v}_n^*$ for $n=1 \text{ to } N+1$, such that the projected representations ($\v_n^{*\top} \x_n$) are highly correlated to each other.}

%Both the CCA and multi-CCA are based on deriving linear transforms on the {\color{black} data-views}. In the next section, we describe the proposed model of deriving non-linear transforms from the {\color{black} data-views} using deep learning. 

\section{Proposed Analysis Approach}\label{sec:deepCCA}

\subsection{Deep CCA}
A deep CCA model finds a pair of optimal non-linear transforms for the two views of the dataset through a pair of neural  networks  such that the two new projections are highly correlated~\cite{andrew2013deep}. As before, let random vectors $ \mathbf{x} \in \mathcal{R}^{\mathcal{D}_1}$ and $\mathbf{y} \in \mathcal{R}^{\mathcal{D}_2}$, denote the data-views. Let $f_1(\cdot)$ and $f_2(\cdot)$ denote the  non-linear functions realized by the neural networks operating on $\mathbf{x}$ and $\mathbf{y}$ respectively. Let $\boldsymbol{\theta}_1$ and $\boldsymbol{\theta}_2$ represent the trainable parameters of $f_1$ and $f_2$ respectively. We find their optimal values by solving the following optimization problem: 
\begin{equation}\label{eqn:dccaCost}
\left(\boldsymbol{\theta}_1^*, \boldsymbol{\theta}_2^*\right)=\underset{\left(\boldsymbol{\theta}_1, \boldsymbol{\theta}_2\right)}{\operatorname{argmax}} \  \rho \left(f_{1}\left(\mathbf{x} ; \boldsymbol{\theta}_1\right), f_{2}\left(\mathbf{y} ; \boldsymbol{\theta}_2\right)\right)
\end{equation}
where $\rho$ corresponds to the cross correlation coefficient. 
% The deep CCA model is shown in Figure $1$.

For a batch of $m$ examples from each of the ($\x$, $\y$) data-views, let $\mathbf{H_x}  \in \mathbb{R}^{d \times m}$ denote the matrix whose columns are the  neural  network outputs $f_1(\cdot)$. Similarly, let $\mathbf{H_y} \in \mathbb{R}^{d \times m}$ denote the outputs from the second network $f_2(\cdot)$. 
% {\color{black} Figure~\ref{fig:dcca_model} shows the deep CCA model with its inputs, outputs and the training objective.}

% \begin{figure}[t!]
% %   \framebox{\includegraphics[width=8cm]{deep_arch.png}}
%   \includegraphics[width=8.6cm]{deep_arch.png}
%   \centering
%   \caption{The deep CCA model. Here, the left side denotes stimulus processing, while right side denotes transformations on the response. }
%     \label{fig:dcca_model}
% \end{figure}

Let, $\mathbf{\bar{H}_x} = \mathbf{H_x} - \frac{1}{m}\mathbf{H_x}\mathbf{1}$ and similarly, $\mathbf{\bar{H}_y} = \mathbf{H_y} - \frac{1}{m}\mathbf{H_y}\mathbf{1}$ denote the centred data matrices, where $\mathbf{1}$ is an all-$1$ matrix of dimension $m \times m$. The covariance matrices of the feed-forward network outputs are given by,
\begin{align}
{ \mathbf{\hat{C}_{xx}}} = \frac{1}{m-1} \mathbf{\bar{H}_x}\mathbf{\bar{H}_x}^{\top} &~;~
{ \mathbf{\hat{C}_{yy}}} = \frac{1}{m-1} \mathbf{\bar{H}_y}\mathbf{\bar{H}_y}^{\top} \nonumber \\
{ \mathbf{\hat{C}_{xy}}} &= \frac{1}{m-1} \mathbf{\bar{H}_x}\mathbf{\bar{H}_y}^{\top} \nonumber
\end{align}

% \begin{eqnarray}
% \mathbf{C_{H_{x}H_{x}}} = \frac{1}{m}\mathbf{\bar{H}_x}\mathbf{\bar{H}_x}^{\top} ~,~
% \mathbf{C_{H_{y}H_{y}}} = \frac{1}{m}\mathbf{\bar{H}_y}\mathbf{\bar{H}_y}^{\top} \\ \nonumber 
% \mathbf{C_{H_{x}H_{y}}} = \frac{1}{m}\mathbf{\bar{H}_x}\mathbf{\bar{H}_y}^{\top}
% \end{eqnarray}

% In order to ensure that the matrices are covariance matrices are positive definite,  $\mathbf{C_{H_{x}H_{x}}}$ and $\mathbf{C_{H_{y}H_{y}}}$ definitions can be added with $\epsilon \mathbf{I}$.  

% Let, $\boldsymbol{T}$ $\triangleq$ $\mathbf{C_{xx}}^{-1/2}\mathbf{C_{xy}}\mathbf{C_{yy}}^{-1/2}$. It can be shown that the gradient of $\rho\left(\mathbf{H_x}, \mathbf{H_y}\right)$ is given by,
Further, let 
\begin{equation}\label{eqn:TH}
\T_H \triangleq { \hat{\C}_{\x\x}^{-1/2}\hat{\C}_{\x\y}\hat{\C}_{\y\y}^{-1/2} }
\end{equation}
Let $\mathbf{T}_H = \mathbf{U} \mathbf{D} \mathbf{V}^{\top}$ denote the $\text{SVD}$ of $\mathbf{T}_H$. It can be shown that optimization\footnote{\textcolor{black}{A detailed derivation of the gradients is given in Appendix~\ref{sec:appendixA}}}  of Equation~(\ref{eqn:dccaCost}) is given by,
\begin{align}
\frac{\partial \rho\left(\mathbf{H_x}, \mathbf{H_y}\right)}{\partial \mathbf{H_x}}=\frac{1}{m-1}\left(2 \mathbf{\nabla_{\mathbf{x}\mathbf{x}}} \mathbf{\bar{H}_x}+\mathbf{\nabla_{\mathbf{x}\mathbf{y}}} \mathbf{\bar{H}_y}\right)
\label{grads}
\end{align}
where
\begin{align}
\mathbf{\nabla_{\mathbf{x}\mathbf{y}}}={\mathbf{\hat{C}_{xx}}}^{-1 / 2} \mathbf{U V}^{\top} {\mathbf{\hat{C}_{yy}}}^{-1 / 2} \label{eq:delxy} \\
\mathbf{\nabla_{\mathbf{x}\mathbf{x}}}=-\frac{1}{2} {\mathbf{\hat{C}}_{\x\x}}^{-1 / 2} \mathbf{U D U}^{\top} {\mathbf{\hat{C}}_{\x\x}}^{-1 / 2} \label{eq:delxx}
\end{align}

Similar expression can be obtained for gradient with respect to $\mathbf{H_y}$. These gradients are backpropagated to learn the optimal model parameters $\boldsymbol{\theta_1}$ and $\boldsymbol{\theta_2}$. Note that the gradient ascent, for $j=1,2$, is performed as 
\begin{equation}\label{eqn:update}
\boldsymbol{\theta_j}^{t+1} = \boldsymbol{\theta_j}^{t} + \eta \frac{\partial \rho\left(\mathbf{H_x}, \mathbf{H_y}\right)}{\partial  \boldsymbol{\theta_j}^{t}}
\end{equation}
{\color{black} where the $\eta$ is the learning rate for the gradient ascent.}

\begin{figure}[t!]
%   \framebox{\includegraphics[width=8.6cm]{m_merged-2.png}}
  \includegraphics[width=8.6cm]{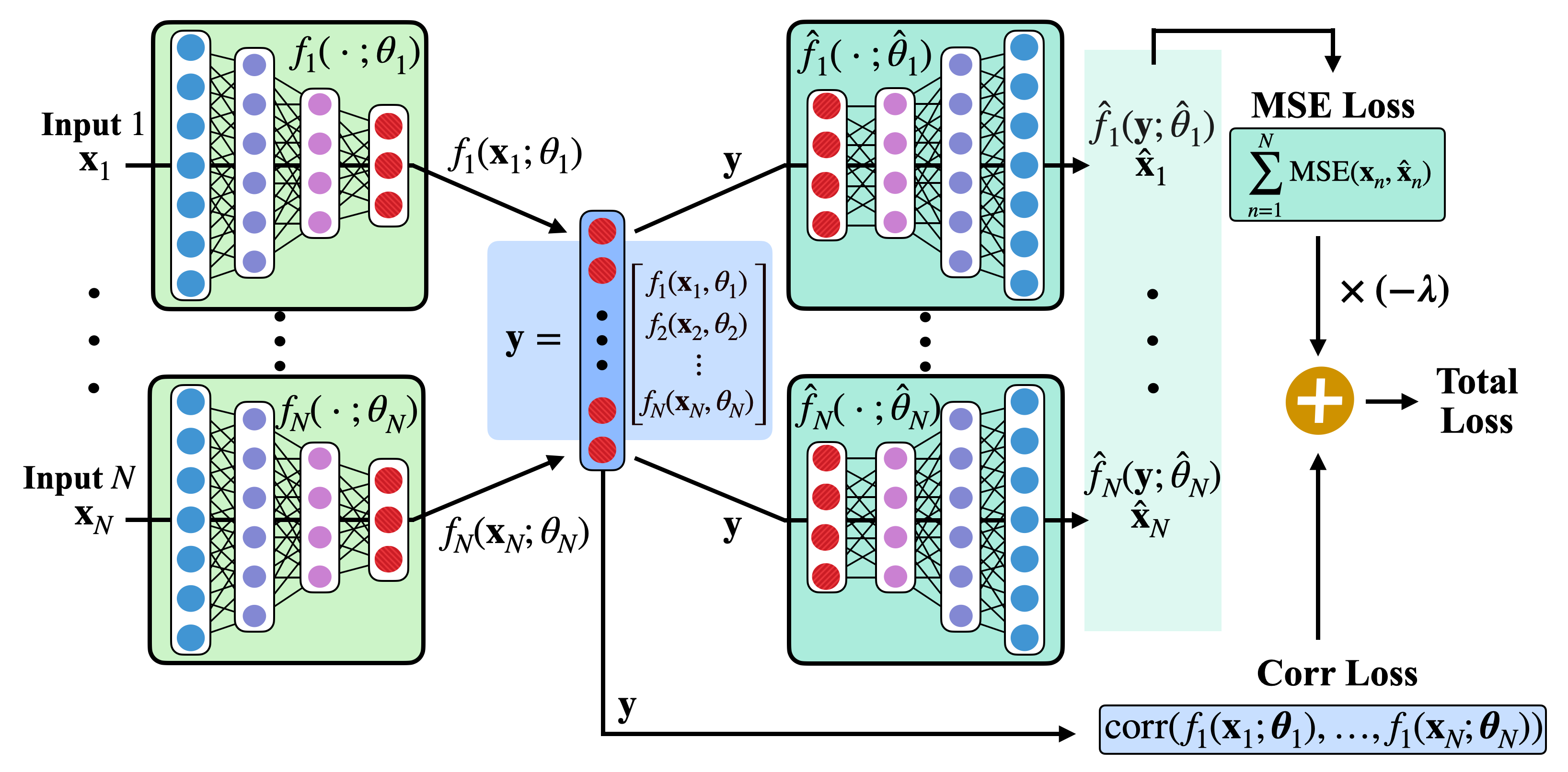}
  \centering
  \caption{The proposed deep multiway CCA Model. $N$ inputs are provided to $N$ encoders. All $N$ outputs are provided to the correlation loss and all $N$ decoders. The decoders' outputs are provided to the reconstruction (MSE) loss. The sum of the correlation loss and negative of the reconstruction loss becomes the cost function.}
    \label{fig:dmcca_model}
\end{figure}

\subsection{Deep Multiway CCA}\label{subsec: DMCCA}

% The deep version of multi-way CCA~\cite{benton2017deep}, attempts to perform multi-way CCA without the linearity assumptions of Sec.~\ref{sec:mcca}. 
For each of the $N$ data-views  $\{\mathbf{x}_n \in \mathcal{R}^{d_n}\}_{n=1}^{N}$, the goal of deep MCCA is to derive optimal non-linear transforms such that the transformed vectors are highly correlated. Let $f_n(\cdot)$, for $n = 1 \text{ to } N$,  represent neural networks with trainable parameters $\boldsymbol{\theta}_n$ operating on $\mathbf{x}_n$.

The $N$ neural networks are trained to maximize the  inter-set correlations defined as:  
\begin{align}
  \rho_{\text{Total}} = \sum_{j=1}^{N} \sum_{k=1, k \neq j}^{N} \rho \left(f_{j}\left(\mathbf{x}_j ; \boldsymbol{\theta}_j\right), f_{k}\left(\mathbf{x}_k ; \boldsymbol{\theta}_k\right)\right)
    \label{rho}
\end{align}

Comparing with Equation~(\ref{eq:isc}), the correlation cost here {\color{black} ($\rho_{\text{Total}}$)} is the summation of pairwise correlation coefficients. {\color{black} While the two definitions are related, the cost based on sum of pairwise correlations is more suitable for gradient based optimization.} The parameters are obtained as:
\begin{align}
    \left(\boldsymbol{\theta}_1^*, \dots, \boldsymbol{\theta}_N^*\right) = \underset{\left(\boldsymbol{\theta}_1, \dots, \boldsymbol{\theta}_N\right)}{\operatorname{argmax}} \rho_{\text{Total}}(\boldsymbol{\theta}_1, \dots, \boldsymbol{\theta}_N)
\end{align}

The backpropagation for each network is similar to the deep CCA model, as described in Equation~(\ref{grads}).

% We have tried further extending this deep version of MCCA.

% \subsubsection{Proposed Model (DMCCA)}
% This version is referred as DMCCA hereafter. 
The proposed model (shown in Figure~\ref{fig:dmcca_model}) has multiple autoencoders sharing encoded representations ($N$ autoencoders for $N$ dataviews respectively). Each view $\mathbf{x}_n$ forward propagates through the encoder part of its autoencoder, $f_n(\boldsymbol{\theta}_n, \cdot)$. All the encoded representations, $f_n(\mathbf{x}_n;\boldsymbol{\theta}_n)$ are concatenated (denoted as $\mathbf{y}$), and propagated to the decoders, $\hat{f}_n(\hat{\boldsymbol{\theta}}_n, \cdot)$.  This shared encoder-decoder model allows the learning of data-view specific transforms that align the views. 
% The proposed model is presented as DMCCA in Fig.~\ref{fig:dmcca_model}.

The model is trained to maximize the joint cost function of correlation and negative of the mean square error (MSE) in reconstruction. This cost function is given as,
\begin{align} \label{eq:dmcca_cost}
    E = \rho_{\text{Total}} - \lambda \sum_{n=1}^{N}\operatorname{MSE}\left(\mathbf{x}_n, \hat{f}_n(\mathbf{y};\hat{\boldsymbol{\theta}}_n)\right)
\end{align}
where $\rho_\text{\color{black} Total}$ is defined by Equation~\eqref{rho} and $\operatorname{MSE}(\cdot)$ is the average squared reconstruction loss. The parameter $\lambda$ controls the trade-off between maximizing the correlation metric and minimizing the MSE in learning the model parameters. 
% The optimum parameters of the autoencoders are obtained by maximizing the cost function of the model, as shown in Equation~(\ref{eq:dmcca_cost}).
% \begin{align}
%     \left(\boldsymbol{\theta}_1^*, \dots \boldsymbol{\theta}_N^* \right) = \underset{\left(\boldsymbol{\theta}_1^*, \dots \boldsymbol{\theta}_N^*\right)}{\operatorname{argmax}} \rho'( \mathbf{x}_1, \dots, \mathbf{x}_N; \boldsymbol{\theta}_1^*, \dots \boldsymbol{\theta}_N^*)
% \end{align}

The model is trained using the $N$ data-views  $\{\mathbf{x}_n\}_{n=1}^{N}$ with the cost metric defined. Note that the correlation loss is independent of the decoder parameters $\boldsymbol{\hat{\theta}_n}$  while the $\operatorname{MSE}(\cdot)$ is a function of both the encoder parameters $\boldsymbol{\theta_n}$ and decoder parameters $\boldsymbol{\hat{\theta}_n}$. Once the model is trained, each data-view $\mathbf{x}_n$ is projected using the encoder $f_n(\mathbf{x}_n;\boldsymbol{\theta}_n)$.
% After training the autoencoders, the encoders' outputs are taken as the denoised outputs. Compared to the DGCCA model, the additional decoders are used for incorporating the $\operatorname{MSE}$ regularization to the correlation loss. It is found that the ISC, among the encoders' outputs, improves in the presence of $\operatorname{MSE}$ regularization. The DGCCA can be viewed as a variant of DMCCA model with $\lambda = 0$.

% All the further results presented in this paper are obtained using the DMCCA model.

\section{Audio-EEG Setup}
\label{sec:Setup}
\subsection{Datasets}
\label{sec:EEGdata} \label{subsec:preprocess}

We experiment our methods on two datasets. The first one is a dataset of EEG responses for speech stimuli, recorded by Liberto et al. \cite{di2015low}. The second dataset is NMED-H~\cite{kaneshiro2016naturalistic}. It contains EEG recordings for a music listening task.

{\textbf{Speech-EEG dataset}}: The EEG recordings were collected using $128$ channels, when the subjects were listening to a male speaker reading snippets of a novel\footnote{\color{black} The speech-EEG dataset is available open source at https://datadryad.org/stash/dataset/doi:10.5061/dryad.070jc}. A Biosemi system, sampled at $512$ Hz, was used to collect the EEG data. We perform the same preprocessing steps as described in Cheveigné et al.~\cite{de2018decoding}. Specifically, the EEG data are down-sampled to $64$ Hz and processed using noise suppression software~\cite{de2018robust}. A band-pass filtering with a passband in the range of $0.1-12$ Hz is applied to the EEG data. At the stimulus side,  the speech envelopes sampled at $44,100$ Hz, are squared and smoothed by a convolution with a square window. Finally, the stimuli data are downsampled to $64$ Hz with a cubic-root compression. 
% All experiments are performed on these preprocessed $1$D speech envelope and $128$D EEG recordings.

% \begin{figure}[t!]
% %   \framebox{\includegraphics[width=8.6cm]{lmdcca.png}}
%   \includegraphics[width=8.8cm]{lmdcca_with_stim.png}
%   \centering
%   \caption{LMCCA and DMCCA models used for inter-subject EEG analysis. Here, $\text{D}_1$ to $\text{D}_N$ are the linear transforms for $N$ subjects respectively. $f_1(\cdot)$ to $f_N(\cdot)$ are the non-linear transforms for $N$ subjects respectively. $\text{D}_s$, $f_s(\cdot)$ are the linear and non-linear transforms for the $d_s$ time-lagged stimulus respectively.}
%     \label{fig:ldmcca}
% \end{figure}

\textbf{Music-EEG dataset} : The NMED-H~{\color{black} \cite{KANESHIRO2020116559}} is an open source dataset containing EEG responses to naturalistic music - $4$ versions (normal, time-reversed, phase-scrambled, and shuffled) of $4$ full-length ``Bollywood'' songs, each approximately of $4.5$ minutes long. {\color{black} The last three stimuli versions were chosen to manipulate the temporal features at varying degrees, while the aggregate frequency content of each stimulus was same. The shuffled version imposes  minimal temporal disruption whereas the phase-scrambled versions were considerably distorted~\cite{KANESHIRO2020116559}}.
% extremely. They were developed to analyse their corresponding effect on the final correlation values

In the phase-scrambled subset, three stimuli files were not considered in the analysis as the features had discontinuities. The EEG recordings were recorded from $125$ electrodes at $1$ kHz. Each recording is filtered between 0.3-50 Hz 
% using EGI Net Station zero-phase filters
and downsampled to 125 Hz. 
% The dataset has three differently preprocessed EEG recordings. 
We use the ``Clean EEG" recordings which are cleaned and aggregated on a per-stimulus, per-listen basis. More details on data acquisition and preprocessing are given in   Kaneshiro~\cite{kaneshiro2016toward}{\color{black}\cite{KANESHIRO2020116559}}. 

The stimuli features are extracted as described in Gang et al.\cite{gang2017decoding}. The acoustic features are extracted using the music information retrieval (MIR) toolbox, Version $1.7.2$~\cite{lartillot2007matlab}.
From the stimuli, $20$ features are extracted in $25$ms analysis windows with a $50$\% overlap between frames~\cite{tzanetakis2002musical, alluri2012large}.  {\color{black} The $20$ features are discussed in the Section IX-B as an appendix.} The principal component analysis (PCA) is performed to obtain a $1$D representation (PC$1$) on these $20$ extracted features. The two individual features, root mean square (RMS) and spectral flux, along with the PC$1$ are chosen to obtain a $3$D representation for the stimuli. 
% At the response side, 
The EEG responses are re-sampled to the sampling rate of the acoustic features ($80$ Hz). 
% All experiments are performed on these preprocessed $3$D features plus $1$D envelope of stimuli and the $125$D clean-EEG recordings.

\subsection{CCA Methods} \label{subsec:Audio-EEG CCA models}
In all our experiments, the linear CCA (LCCA)~\cite{de2018decoding} and linear MCCA (LMCCA)~\cite{de2019multiway} analysis act as the baseline setup for comparison with the deep CCA (DCCA) and the deep multiway CCA (DMCCA) methods. 
For the multi-subject EEG analysis, the outputs from the multiway CCA are further processed with CCA (either LCCA or DCCA).

\begin{figure}[t!]
  \includegraphics[width=8.6cm]{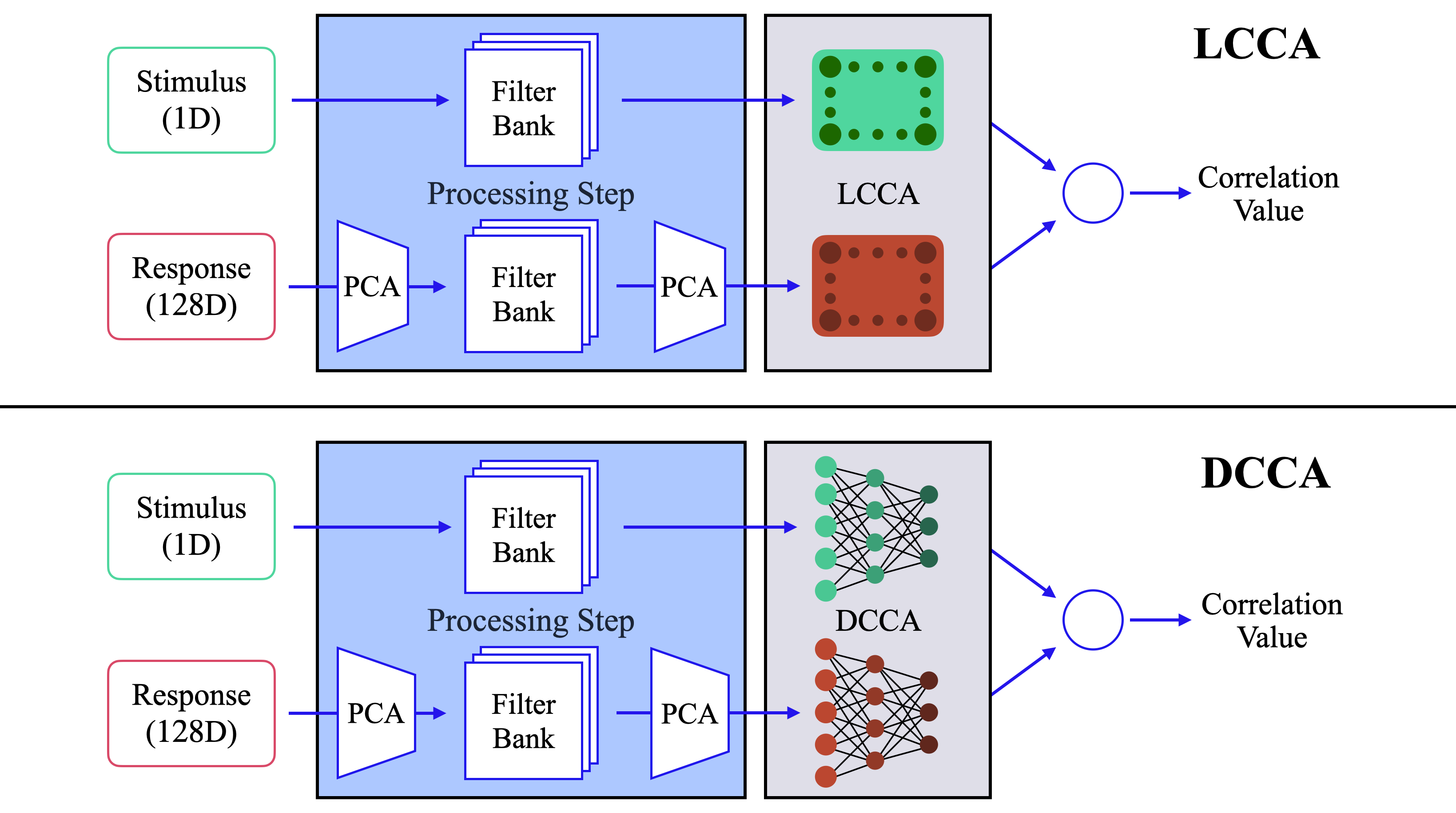}
  \centering
  \caption{The LCCA and DCCA methods. The stimulus features are low-pass filtered envelopes.}
    \label{fig:ldcca}
\end{figure}

\subsubsection{LCCA}
% It has been shown that the linear CCA model successfully provides better correlation values compared to other linear models~\cite{de2018decoding}. The LCCA method is performed on each subject separately. 
On the $1$D preprocessed stimuli data, a dyadic bank of $21$ FIR band-pass filters is applied that contains filters that are  approximately uniformly distributed on a logarithmic scale~\cite{de2018decoding}. At the response end, a PCA is applied to the EEG data that transforms the $128$D (or $125$D for music data) EEG data to $60$D. The filterbank is applied on these $60$D EEG data to yield $1260$D data. A second PCA is applied that transforms them to $139$D subspace. Now, the $21$D stimuli data and the $139$D EEG data are projected onto common subspace using CCA transforms.
{\color{black} The data are processed (the choice of PCA and the dimensions after each stage) as proposed by Cheveigné et al.~\cite{de2018decoding}. The combination of PCA and filterbank acts as a spatio-temporal filter on the data. }

\subsubsection{DCCA} 
The neural networks used in DCCA model have a $2$ hidden layer architecture, for each view, with $2038$ and $1608$ units for the first and second layers respectively followed by a $d$ dimensional output layer. The data are processed similar to the LCCA method, and the final $21$D and $139$D representations are input to a deep CCA model.
Figure~\ref{fig:ldcca} describes the LCCA and DCCA methods. 
% Both the LCCA and DCCA methods follow the same processing steps on the stimuli-responses data.
% from Section~\ref{subsec:preprocess}. 

% \subsubsection{LMCCA}
% The linear MCCA model is used as a denoising step for EEG data of multiple subjects. EEG response data, for common stimuli, of $N$ subjects are aggregated. It filters out components that are not common among the response data of multiple subjects. 
% % As the response data is collected for same stimulus, we can assume that the components which are not common among various subjects are subject-specific and unrelated to the provided stimulus. 
% From each subject's response data, it tries to extract $d$ signals that are common across all the subjects' response data~\cite{de2019multiway}.
% % It is performed as proposed by de Cheveigné et al.~\cite{de2019multiway}.

% For each subject's response data, the linear MCCA model provides an optimum linear transform that filters out the subject-specific components across the subjects data. $N$ subjects' EEG data, along with the $d_s$D time-lagged common stimulus, are provided to the model which returns $N$ denoised EEG data for each subject.

\subsubsection{LMCCA}
The preprocessed EEG responses from $N$ subjects and a time-lagged version of their common stimuli ($d_s$D), are provided to a linear MCCA model to obtain the denoised representations for each subject's EEG response. Now, each subject's denoised EEG data and their corresponding stimuli can be provided, separately, to the LCCA and DCCA methods to obtain final representations. This is performed as proposed in Cheveigné et al.~\cite{de2019multiway}.

\subsubsection{DMCCA}
The preprocessed EEG responses, along with the common stimuli are provided to a deep MCCA model to obtain a $d$D denoised representation for each EEG response. 

The architecture of the DMCCA model is shown in Figure~\ref{fig:dmcca_model}. The encoder has two hidden layers of $60$ units each and an output layer of $d$ units. The decoding part has two hidden layers of $60$ and $110$ units respectively. 
% We also use the same stimuli features in the DMCCA model as the  LMCCA model as shown in Figure~\ref{fig:ldmcca}. 
% {\color{black} Figure~\ref{fig:ldmcca} shows how both the MCCA methods take the $N+1$ inputs and obtain a linear or non-linear transform for each input.}

The $d_s$ and $d$ are hyperparameters and the best values are selected by varying them, for both the variants of MCCA. More details are discussed in Section~\ref{subsec:hyper}. The outputs from the linear MCCA model are of $128$D for speech ($125$D for music) dataset. For both the MCCA methods, the denoised responses are provided to the filterbank followed by a PCA to generate $139$D vectors. The $d$D stimuli obtained are projected onto a $1$D subspace using PCA, followed by the filterbank resulting in a $21$D data. These steps make sure that the inputs to the CCA transforms are equivalent in both versions of MCCA. 
%This process of filtering, performing PCA and then CCA is equivalent to the CCA3 models performed after LMCCA. 
% Fig.~\ref{fig:ldmldc3} shows how the denoised data from LMCCA and DMCCA models are provided to the CCA3 models (or their equivalents). 

% These denoised $128$D EEG data, from linear or deep MCCA model, and the preprocessed $1$D stimulus data can be provided to the linear and deep CCA models, which obtain the final representations. The denoising transforms from the MCCA models and the projecting transforms from the CCA models are obtained using the training data. These transforms are then applied on test data to obtain their final representations. The correlation value of these final representations is used to measure and compare these models.

\begin{figure}[t!]
  % \framebox{\includegraphics[width=8.6cm]{lmdcca3.png}}
  \includegraphics[width=8.8cm]{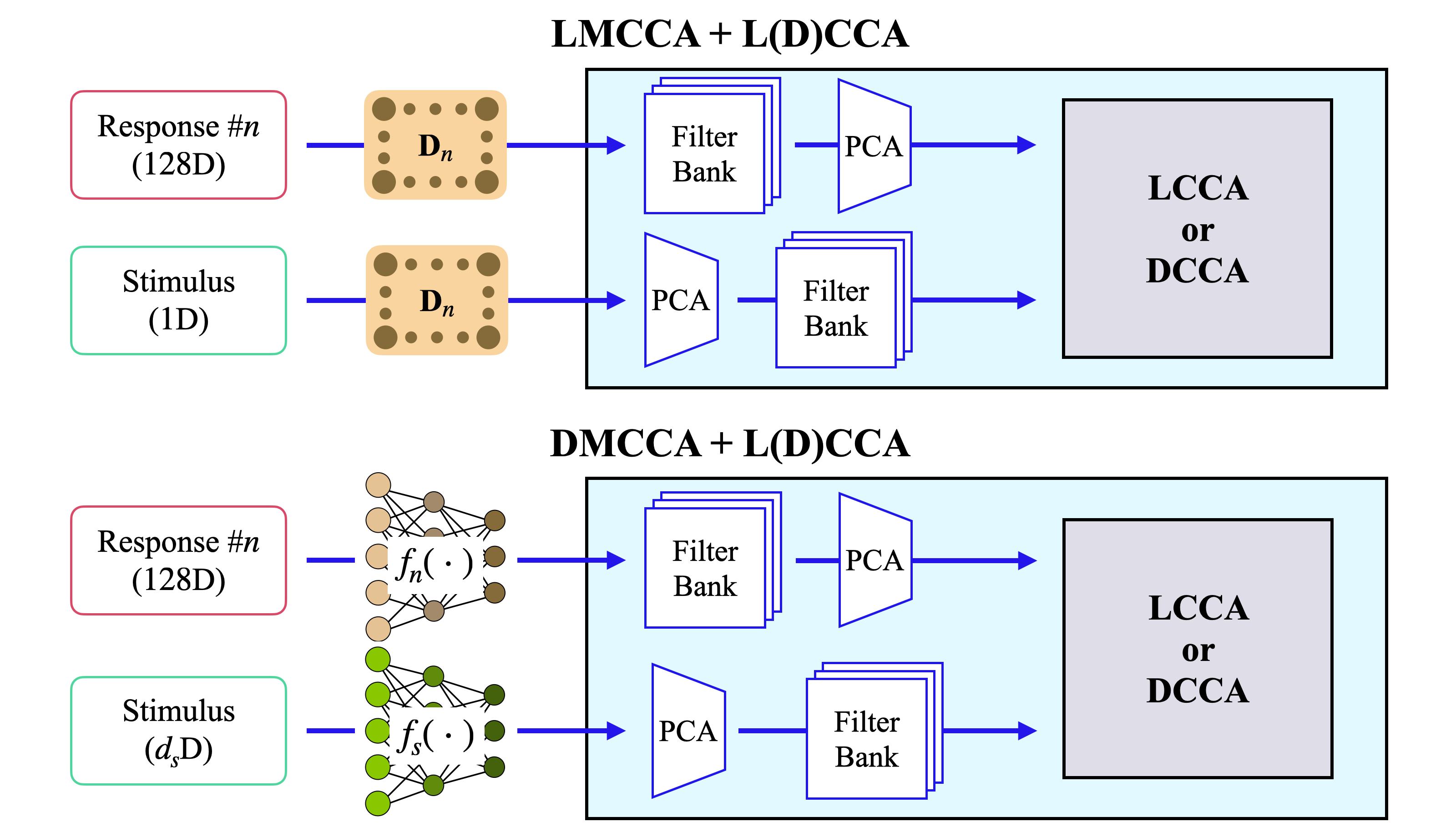}
  \centering
  \caption{Configuration of the four systems - linear multiway CCA with linear CCA (LMLC), linear multiway CCA with deep CCA (LMDC), deep multiway CCA with linear CCA (DMLC) and deep multiway CCA with deep CCA (DMDC) methods. }
   \label{fig:ldmldc3}
\end{figure}

For intra-subject analysis, the LCCA/DCCA are performed on each subject's data separately. For inter-subject analysis, the LMCCA/DMCCA are performed on data from multiple subjects data followed by a subject-specific LCCA/DCCA method. 
% before generating the correlation score. 
Thus, we have four combinations 
\begin{enumerate}
  \item \textbf{LMLC}: LMCCA + LCCA 
  \item \textbf{LMDC}: LMCCA + DCCA 
  \item \textbf{DMLC}: DMCCA + LCCA 
  \item \textbf{DMDC}: DMCCA + DCCA 
\end{enumerate}

Figure~\ref{fig:ldmldc3} shows the four combinations of the MCCA denoising followed by CCA analysis for each subject.

% To summarize, $6$ methods are tried on the two datasets. One set of experiments compares the linear and deep versions of CCA$3$ method. Second set of experiments, LMLC and LMDC, denoises EEG responses using linear MCCA and obtain final representations for each subject using both versions of CCA$3$ models. The third set replaces the linear MCCA with DMCCA model, resulting in DMLC and DMDC. 
% % Here denoising is taken in the context of removing the signals unrelated to the common stimulus. 
\subsection{Experimental setup}\label{sec:exptSetup}
% All the methods get the preprocessed data from Section~\ref{subsec:preprocess} as inputs. 
For the speech dataset, the methods are performed on the preprocessed $1$D stimuli envelopes and $128$D EEG responses. For NMED-H, along with the $1$D stimuli envelopes, each dimension of the $3$D preprocessed stimuli is also used with the $125$D clean EEG recordings.

From the speech dataset, stimulus-response data of  $8$ subjects are considered  randomly to perform the experiments. The data from each subject contains $20$ sessions with approximately $160$s of recording in each session. For all the methods, $20$ fold validation experiments are performed where $18$ sessions are used for training, one session for validation and the remaining session for testing. Given a sampling rate of $64$ Hz, the approximate number of instances for the linear/deep model training per subject is about $200$k.
% The data are preprocessed as discussed earlier in section \ref{subsec:preprocess}.

{\color{black} The NMED-H dataset contains recordings from $48$ subjects and $16$ stimuli. The subjects were divided into $16$  groups of $12$ subjects with each subject appearing in $4$ groups. Each group is presented with $2$ trials of a stimulus which results in each subject listening to $2$ trials of $4$ different stimuli.  In our analysis, we have used all the data available. All the $12$ subjects that were available for each stimulus were used in the inter-subject analyses and the intra-subject analysis.}
% and the average correlation coefficient, $\rho$, (for each stimuli version and stimuli feature) of the final representations are compared.}

We split the data into $90-5-5$ for training, validation and test respectively. It results in about  $155$k samples for training and $8.5$k samples for testing and validation, for each subject in the CCA experiments. Similarly, we use  
$38$k samples for training and $2$k samples for testing and validation, for each subject per stimulus  in the MCCA experiments. 
% The MCCA methods are performed for $12$ subjects' EEG data along with the common stimulus.

A leaky ReLU activation function with a negative slope coefficient of $0.1$ is used in the DCCA model and the encoder part of the  DMCCA model. A linear activation function is used at the output layer of the decoder sections in the DMCCA model\footnote{The implementation is available at   https://github.com/iiscleap/deep-cca-for-audio-EEG}. Further, dropout regularization~\cite{hinton2012improving,srivastava2014dropout} is incorporated in the deep models training to avoid over-fitting in the noisy conditions.

\begin{figure*}
  \centering
  \includegraphics[width=18cm]{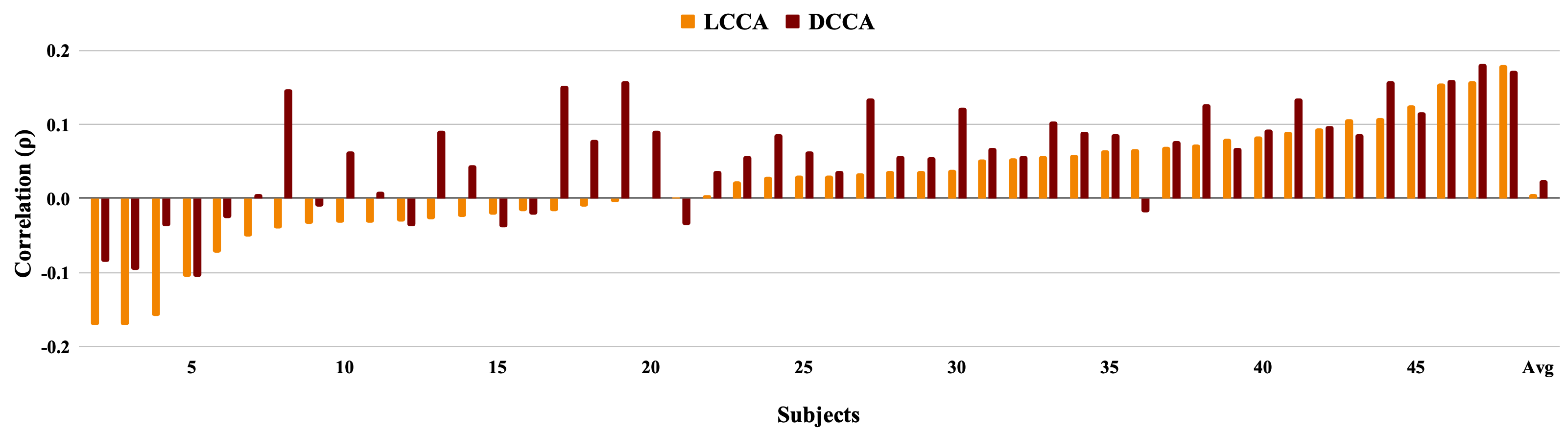}
    \label{fig:lvsdcca3_nmedh}
  \caption{Comparison between LCCA and DCCA methods for PC$1$ stimuli features, of the $48$ subjects from the NMED-H dataset, arranged in the increasing order of the LCCA correlation values. The last column shows the average of the $48$ subjects.}
\end{figure*}

% To summarize, six methods, three pairs of methods, have been tried on the data. One set of experiments compares the already established linear CCA$3$ with the proposed DCCA$3$ method. Another set of experiments, talks about the benefits of denoising each subject's response using linear or deep MCCA method. Here denoising is taken in the context of removing the signals unrelated to the common stimulus. So, the assumption is that the MCCA step identifies the stimulus related components in each response and results a transform which removes all the components unrelated to the stimulus. Because, if a signal is not common across the subjects, it might be a subject-specific signal rather than a stimulus driven one. And the MCCA part is where the communication across the subjects happens. Another set of experiments tries the linear and deep CCA$3$ method on all subjects' response stacked for a given stimulus. It just talks about how the CCA$3$ results vary by stacking up other subjects data. 
{\textbf{Performance Metric}}:
 The primary metric used is the Pearson correlation between the transformed EEG and audio signals on the held-out test set.
{\color{black} For each subject, the LCCA and DCCA methods performance is measured by the correlation coefficient ($\rho$) of the two final representations. The LMCCA tries to maximize the $\rho_\text{ISC}$, and the DMCCA tries to maximize the $\rho_\text{Total}$. 
%The four CCA combinations' performance is measured as the average correlation coefficient ($\rho$) of the respective subjects' LCCA and DCCA final representations. 
For overall results, instead of direct averaging of Pearson correlation values which is mathematically incorrect, we perform a  $z$-score based averaging implemented in the Statsoft software  \cite{statsoft2001statistica}.}

We also use a secondary performance metric based on classification of aligned versus misaligned EEG-audio segments \cite{de2018decoding}. Here, fixed-length segments of EEG and audio signals that are randomly located are correlated using the models. If the audio and EEG segments are aligned, the model is expected to generate a higher correlation score than when the two signals are misaligned. The correlation scores are analyzed using the sensitivity index (Cohen's $d'$ statistic).     

{\color{black} Let the means of the matched and mismatched segments' correlation coefficients be $\mu_1$ and $\mu_2$ respectively. Let, $\sigma_1^2$ and $\sigma_2^2$ be their respective variances. The Cohen's $d'$ is:  
\begin{equation}
    d^{\prime} = \frac{|\mu_1 - \mu_2|} {\sqrt{\frac{1}{2} \left(\sigma_1^2 + \sigma_2^2 \right)}}
\end{equation}}
\section{Results}
\label{sec:results}
% For the speech dataset, stimuli-response data for $6$ subjects are taken. For the music-EEG dataset, the data from $48$ subjects are used. 
The results comparing the linear and deep CCA models for intra-subject and inter-subject experiments on the speech-EEG dataset are given in Table~\ref{tab:lccaVSdcca} and Table~\ref{tab:lmccaVSdmcca} respectively. The results for the music-EEG dataset are shown in Table ~\ref{tab:music_lcca} and Table ~\ref{tab:music_mcca}. Pairwise one-tailed t-tests are performed on the pairs of LCCA-DCCA (Table I and III) and LMLC-DMDC (Table II and IV). % The corresponding $p$ and $t$-values are shown in each table.

\begin{table}
\caption{pearson correlation values for { lcca} and  { dcca} methods on the speech - { eeg } dataset.  {a} paired t-test with  \{p-value\} and [t-value] are indicated. \textcolor{black}{ statistically significant ($p < 0.025$) for $4$ out of $8$ subjects and the overall aggregate.}}
\label{tab:lccaVSdcca}
\begin{center}
\begin{tabular}{|c|cc|c| }
    \hline 
    Models & LCCA & DCCA & t-test\\
    \hline
    Sub1  & 0.220 & \textbf{0.275} &{\fontsize{8}{10} \selectfont \{0.01\}[2.3]} \\
    Sub2  & 0.258 & \textbf{0.316}&{\fontsize{8}{10} \selectfont \{0.01\}[2.5]} \\
    Sub3  & 0.175 & \textbf{0.213}&{\fontsize{8}{10} \selectfont \{0.051\}[1.7]} \\
    Sub4  & 0.316 & \textbf{0.403}&{\fontsize{8}{10} \selectfont \{1e-3\}[3.3]} \\
    Sub5  & 0.307 & \textbf{0.338}&{\fontsize{8}{10} \selectfont \{0.03\}[1.9]} \\
    Sub6  & 0.315 & \textbf{0.354}&{\fontsize{8}{10} \selectfont \{0.049\}[1.7]} \\
    \color{black}Sub7  & \color{black}0.260 & \color{black}\textbf{0.292}&\color{black}{\fontsize{8}{10} \selectfont \{0.06\}[1.5]} \\
    \color{black}Sub8  & \color{black}0.183 & \color{black}\textbf{0.232}&\color{black}{\fontsize{8}{10} \selectfont \{0.02\}[2.2]}\\
    \hline
    Overall  & \color{black}0.255 & \color{black}\textbf{0.304} & 
    \color{black}{\fontsize{8}{10} \selectfont \{1e-6\}[4.8]} \\
    \hline
\end{tabular} 
\end{center}
\end{table}

\begin{table}[t!]
\caption{pearson correlation values for { lmlc },  { lmdc}, { dmlc} and { dmdc}.  { a }statistical significance test (t-test) between  {lmlc} and  {dmdc} ( \{p-value\}[t-value]) is also reported. \textcolor{black}{the improvements are statistically significant ($p < 0.025$) for $5$ out of $8$ subjects and the overall aggregate.}}
\label{tab:lmccaVSdmcca}
\begin{center}
\begin{tabular}{|c|cccc|c|}
    \hline
    Models & LMLC & LMDC & DMLC & DMDC & t-test \\ \hline 
    Sub1 & 0.262 & 0.271 & 0.375 & \textbf{0.377}&{\fontsize{8}{10} \selectfont \{9.1e-7\}[5.6]} \\
    Sub2 & 0.289 & 0.325 & 0.367 & \textbf{0.374}&{\fontsize{8}{10} \selectfont \{5.1e-4\}[3.6]} \\ 
    Sub3 & 0.160 & 0.177 & 0.258 & \textbf{0.259}&{\fontsize{8}{10} \selectfont \{6.3e-5\}[4.2]} \\
    Sub4 & 0.310 & \textbf{0.378} & 0.341 & 0.361&{\fontsize{8}{10} \selectfont \{3.6e-2\}[1.8]} \\ 
    Sub5 & 0.309 & 0.354 & 0.389 & \textbf{0.392}&{\fontsize{8}{10} \selectfont \{8.5e-5\}[4.2]} \\ 
    Sub6 & 0.327 & 0.342 & 0.416 & \textbf{0.420}&{\fontsize{8}{10} \selectfont \{4.6e-5\}[4.4]} \\ 
    \color{black}Sub7 & \color{black}0.275 & \color{black}0.289 & \color{black}\textbf{0.310} & \color{black}\textbf{0.310}&\color{black}{\fontsize{8}{10} \selectfont \{4.4e-2\}[1.7]}\\ 
    \color{black}Sub8 & \color{black}0.221 & \color{black}0.245 & \color{black}0.259 & \color{black}\textbf{0.272}&\color{black}{\fontsize{8}{10} \selectfont \{2.8e-2\}[2.0]} \\ 
    \hline
    % Overall & \color{black}0.276 & \color{black}0.308 & \color{black}0.358 & \color{black}\textbf{0.364}&\color{black}{\fontsize{8}{10} \selectfont \{9e-14\}[7.7]} \\ 
    Overall & \color{black}0.270 & \color{black}0.299 & \color{black}0.339 & \color{black}\textbf{0.344}& {\color{black}{\fontsize{8}{10} \selectfont \{9e-14\}[7.7]}} \\ 
    \hline
\end{tabular}
\end{center}
\vspace{-0.4cm} 
\end{table}

\begin{table*}[ht!]
    \caption{{average} correlation values for $48$ subjects from the { nmed-h} dataset in intra-subject analysis. a statistical significance test (t-test) between { lcca} and { dcca} methods (indicated as \{p-value\}[t-value]) is also reported. \textcolor{black}{the improvements are statistically significant ($p < 0.025$) for all experiments.}}
    \label{tab:musicEEGDataset}
    \begin{center}
    
    \label{tab:music_lcca}
    \begin{tabular}{ |c|cc|cc|cc|cc| }
        \hline
        Stimulus feature & \multicolumn{2}{c|}{Normal} & \multicolumn{2}{c|}{Reversed} & \multicolumn{2}{c|}{Phase-Scrambled} & \multicolumn{2}{c|}{Shuffled} \\
        \hline
        CCA Model & LCCA & DCCA \hspace{0.2cm} [t-test] \hspace{0.1cm} & LCCA & DCCA \hspace{0.2cm} [t-test] \hspace{0.1cm} & LCCA & DCCA \hspace{0.2cm} [t-test] \hspace{0.1cm} & LCCA & DCCA \hspace{0.2cm} [t-test] \hspace{0.1cm} \\
        \hline
        Envelope       &  0.007 & \textbf{0.118} {\fontsize{8}{10} \selectfont \{1e-4\}[3.7]} & -0.003 & \textbf{0.117} {\fontsize{8}{10} \selectfont \{3e-5\}[4.1]} & -0.052 & \textbf{0.095} {\fontsize{8}{10} \selectfont \{4e-7\}[5.3]} & -0.013 & \textbf{0.134} {\fontsize{8}{10} \selectfont \{3e-5\}[4.2]} \\
        PC1            & -0.020 & \textbf{0.077} {\fontsize{8}{10} \selectfont \{9e-4\}[3.2]} &  0.012 & \textbf{0.105} {\fontsize{8}{10} \selectfont \{1e-3\}[3.0]} & -0.016 & \textbf{0.072} {\fontsize{8}{10} \selectfont \{1e-3\}[3.1]} &  0.030 & \textbf{0.135} {\fontsize{8}{10} \selectfont \{1e-3\}[3.0]} \\
        RMS            & -0.004 & \textbf{0.087} {\fontsize{8}{10} \selectfont \{2e-4\}[3.6]} &  0.008 & \textbf{0.101} {\fontsize{8}{10} \selectfont \{3e-4\}[3.5]} & -0.042 & \textbf{0.091} {\fontsize{8}{10} \selectfont \{4e-7\}[5.2]} & -0.025 & \textbf{0.100} {\fontsize{8}{10} \selectfont \{3e-5\}[4.2]} \\
        Spectral Flux  &  0.008 & \textbf{0.102} {\fontsize{8}{10} \selectfont \{3e-4\}[3.5]} & -0.004 & \textbf{0.113} {\fontsize{8}{10} \selectfont \{9e-6\}[4.5]} & -0.034 & \textbf{0.107} {\fontsize{8}{10} \selectfont \{3e-7\}[5.3]} & 0.005  & \textbf{0.123} {\fontsize{8}{10} \selectfont \{3e-5\}[4.2]} \\
        \hline
    \end{tabular}
    % \end{center}
% \end{table*}

% \begin{table*}[]
    \vspace{0.4cm} 
       \caption{ {average} correlation values for $48$ subjects from the { nmed-h dataset} in inter-subject analysis. { a} statistical significance test (t-test) between { lmlc} and { dmdc} methods (indicated as \{p-value\}[t-value]) is also reported. \textcolor{black}{the improvements are statistically significant ($p < 0.025$) for  all experiments.}}
   \label{tab:music_mcca}
%   \begin{center}
    \begin{tabular}{ |c|cccc|c|cccc|c| }
        \hline
        Stimulus feature & \multicolumn{5}{c|}{Normal} & \multicolumn{5}{c|}{Reversed}\\
        \hline
        MCCA Model & LMLC & LMDC & DMLC & DMDC & t-test & LMLC & LMDC & DMLC & DMDC & t-test\\
        \hline
        Envelope      &  0.076 & 0.146 & 0.344 & \textbf{0.349} & {\fontsize{8}{10} \selectfont \{3e-26\}[14]} & 0.062 & 0.099 & 0.299 & \textbf{0.384} & {\fontsize{8}{10} \selectfont \{1e-37\}[21]} \\
        PC1           & -0.007 & 0.102 & \textbf{0.384} & 0.321 & {\fontsize{8}{10} \selectfont \{1e-26\}[14]} & 0.030 & 0.159 & \textbf{0.360} & 0.323 & {\fontsize{8}{10} \selectfont \{2e-19\}[11]} \\
        RMS           &  0.001 & 0.114 & \textbf{0.341} & 0.246 & {\fontsize{8}{10} \selectfont \{3e-13\}[8.5]}  & 0.042 & 0.135 & \textbf{0.318} & 0.220 & {\fontsize{8}{10} \selectfont \{1e-08\}[6.2]} \\
        Spectral Flux &  0.017 & 0.110 & 0.341 & \textbf{0.343} & {\fontsize{8}{10} \selectfont \{2e-24\}[13]}  & 0.053 & 0.170 & \textbf{0.340} & 0.321 & {\fontsize{8}{10} \selectfont \{1e-16\}[10]}  \\
        \hline
    \end{tabular} 
    
    \vspace{0.2cm} 
    \begin{tabular}{ |c|cccc|c|cccc|c| }
        \hline
        Stimulus feature & \multicolumn{5}{c|}{Phase-Scrambled} &  \multicolumn{5}{c|}{Shuffled} \\
        \hline
        MCCA Model & LMLC & LMDC & DMLC & DMDC & t-test & LMLC & LMDC & DMLC & DMDC & t-test\\
        \hline
        Envelope      & 0.042 & 0.092 & \textbf{0.312} & 0.299 & {\fontsize{8}{10} \selectfont \{8e-26\}[14]} & 0.077  & 0.132 & \textbf{0.341} & 0.333 & {\fontsize{8}{10} \selectfont \{4e-19\}[11]} \\
        PC1           & 0.012 & 0.166 & 0.262 & \textbf{0.389} & {\fontsize{8}{10} \selectfont \{6e-21\}[12]} & 0.051 & 0.149 & 0.345 & \textbf{0.347} & {\fontsize{8}{10} \selectfont \{5e-21\}[12]} \\
        RMS           & 0.020 & 0.108 & 0.176 & \textbf{0.397} & {\fontsize{8}{10} \selectfont \{2e-07\}[7.4]} & 0.051 & 0.156 & 0.327 & \textbf{0.345} & {\fontsize{8}{10} \selectfont \{1e-19\}[11]} \\
        Spectral Flux & 0.038 & 0.207 & 0.340 & \textbf{0.390} & {\fontsize{8}{10} \selectfont \{3e-22\}[12]} & 0.061 & 0.145 & 0.294 & \textbf{0.322} & {\fontsize{8}{10} \selectfont \{8e-15\}[9.2]} \\
        \hline
    \end{tabular}
    \end{center}
    \vspace{-0.5cm}
\end{table*}

\subsection {Speech-EEG dataset results}
For speech-EEG experiments, the $20$ cross-validation results (correlation values) for the $20$ folds for all the subjects are considered for a pairwise t-test. As seen in Table~\ref{tab:lccaVSdcca}, the DCCA improves over the LCCA for all the subjects. The average absolute improvements for DCCA over the LCCA in terms of correlation value is $5$\%.  {\color{black} The improvements are also statistically significant ($p<0.025$)\footnote{\textcolor{black}{Compensating for multiple comparisons on two datasets, the significance threshold ($\alpha=0.05$) is divided by $2$ to obtain a threshold of $0.025$ as per the Bonferroni correction method~\cite{aickin1996adjusting}.}} for  $4$ out of $8$ subjects and the overall aggregate.} 
% For LMLC and DMLC methods also, the improved correlations are found to be statistically significant ($p<0.01$) for $5$ out of $6$ subjects.

The comparison of various inter-subject experiments on the speech-EEG dataset is shown in Table~\ref{tab:lmccaVSdmcca}. \textcolor{black}{Here, the inter-subject alignment using linear method (LMLC) improved over the linear intra-subject correlation model (LCCA) on all subjects except subject 3 and 4. The inter-subject alignment using deep learning (DMLC/DMDC)  improves the correlation scores  compared to the intra-subject scores reported in Table~\ref{tab:lccaVSdcca} for all cases except subject 4.} Further, the deep models consistently improve over the linear counterparts. In particular, the deep multiway CCA approach improves over the linear multiway CCA by an absolute correlation value of {$7.4$} \% on the average. \textcolor{black}{The improvements are  found to be statistically significant ($p<0.025$) for $5$ out of $8$ subjects. The overall aggregate score is found to be statistically significant as well.}

\subsection{Music-EEG dataset results}
For LCCA/DCCA methods, the average correlation values for the $48$ subjects in the NMED-H dataset is reported in Table~\ref{tab:musicEEGDataset}. The results are reported for different music conditions - normal, shuffled, time-reversed and phase-scrambled; and stimuli features - envelope, PC1, RMS and spectral flux. \textcolor{black}{The pair-wise t-test on the NMED-H dataset shows that all improvements obtained by the deep versions are statistically significant ($p<0.025$). The performance of LCCA and DCCA methods on the PC$1$ features of $48$ subjects from the NMED-H dataset is shown in Figure 6. The average absolute improvements are about $11$\% for the DCCA over LCCA method. For inter-subject analysis, the {\color{black} Table~\ref{tab:music_mcca} shows that the} DMDC improves over the LMLC method with an average absolute improvement of $29.3$\%.}

% 0.3440625 - 0.039125 = 0.3049375

%               sub1  |  sub2  |  sub3 | sub4  |  sub5 | sub6    | Combined
% p-values
% CCA        - 0.7005  | 0.2439 |0.5367 |0.0106 |0.0531 |0.5410  | 0.0026
% MCCA        - 1.18e-5| 0.0502 |4.8e-4 |0.0741 |2.0e-4 |4.9e-4  | 1e-27
% Interaction - 1      | 1      |1      |0.9293 |0.9970 |1       | 1
% t-values
% CCA        -  0.1501 | 1.3988 |0.3883 |7.1967 |3.9732 |0.3803  | 9.2433
% MCCA        - 4.8941 | 1.8513 |3.4480 |1.7186 |3.7684 |3.4426  | 5.2177
% Interaction - 0.0699 | 0.1163 |0.0437 |0.5327 |0.2966 |0.0217  | 0.2118

\subsection{Statistical Analysis}

\textcolor{black}{In order to measure the significance of the improved correlations of our proposed deep models over the baseline system, we  have performed two statistical tests: a) one-tailed pairwise t-test and b) Cohen's $d'$. The pairwise t-test is performed to objectively quantify the difference in the distribution of the correlation scores from the two methods. \textcolor{black}{Given that the same hypothesis (LCCA versus DCCA on intra-subject analysis or LMLC versus DMDC on inter-subject analysis) is tested on two different datasets (speech-EEG and music-EEG), a compensation is required for multiple comparisons. We use the Bonferroni  correction~\cite{aickin1996adjusting}. Thus, a p-value  $0.05/2 = 0.025$ is used to check if the improvements in the correlation are statistically significant on each dataset.}}
\color{black} The pairwise t-test results   comparing the linear and deep models are reported for  speech-EEG (Table~\ref{tab:lccaVSdcca},~\ref{tab:lmccaVSdmcca})  and music-EEG (Table~\ref{tab:music_lcca},~\ref{tab:music_mcca}).

% In addition, a two-way ANOVA test on the LMLC, LMDC, DMLC and DMDC methods is performed for each subject. It is performed with the version of MCCA and CCA as the two binary factors.  $1$ out of 6 subjects has shown that 

\begin{figure}
  % \framebox{\includegraphics[width=8.6cm]{lmdcca3.png}}
  \includegraphics[width=8.8cm]{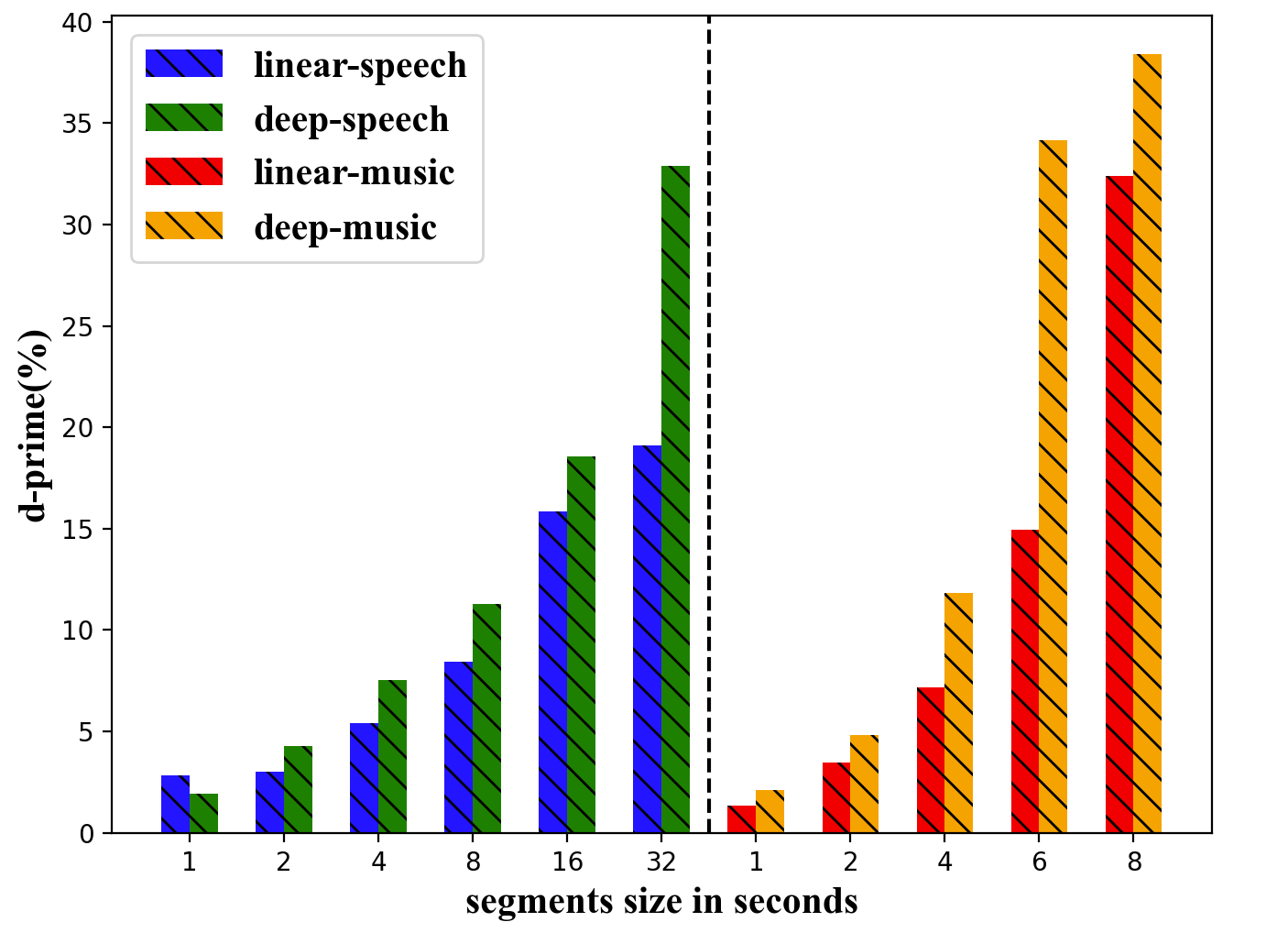}
  \centering
  \caption{The $d'$ metric for both the datasets. linear-speech and linear-music correspond to the $d'$ values of the final representations from LMLC models for  the speech and music datasets. Similarly, deep-speech and deep-music correspond to DMLC models.}
   \label{fig:d-prime}
   \vspace{-0.4cm}
\end{figure}

As mentioned in Section~\ref{sec:exptSetup}, a classification metric is also performed where audio-EEG segments are classified as aligned or misaligned based on the Pearson correlation measure. 
\textcolor{black}{The second statistical test, Cohen's $d'$ \cite{cohen2013statistical}, is an effect size used to indicate the standardised difference between two classes (in our case, these classes are aligned and mis-aligned audio-EEG pairs). The $d'$ metric   quantifies the model's  ability to match the corresponding stimulus-response pair based on the correlation value, $\rho$. The test data is divided into $N$ segments of $t$ seconds each, and the correlation coefficient values $\rho$ are calculated for the linear the deep methods for both aligned and mis-aligned segments (speech/audio-EEG pairs).  The Cohen's $d'$ metric measures the model's ability to separate aligned versus mis-aligned pairs.} 
\color{black}
The LMLC/DMLC methods are used on the audio-EEG segments and the correlation values are computed. Using the correlation score from the respective models,  the Cohen's $d'$ is computed on the correlation score. The $d'$ statistics are presented in Figure~\ref{fig:d-prime}. This is performed separately for the speech-EEG and music-EEG datasets.  As seen in Figure~\ref{fig:d-prime}, the deep model improves over the linear model in all the cases except for $1$ second segments in speech-EEG data. In longer segments, considerable improvements in the $d'$ statistic are observed for the deep models.

% LCCA3, DCCA3: ($0.013, 0.008, 0.050, 0.001, 0.031, 0.052$)
% LMLC, DMLC: ($3.44e-06, 0.002, 9.699e-05, 0.145, 0.0002, 6.212e-05$)

% \begin{figure}[t!]
% %   \framebox{\includegraphics[width=8.6cm]{LMLCvsDMLC.png}}
%   \includegraphics[width=8.8cm]{LMLCvsDMLC.png}
%   \centering
%   \caption{Comparison between LMLC and DMLC for 20 cross-validation sessions of a subject from the first dataset. MCCA models were trained with one more subject from the same dataset.}
%     \label{fig:lvsdmcca}
% \end{figure}

\color{black}
\section{Discussion}
\label{sec:discussion}

% The MCCA methods show that other subjects' response data can be helpful to obtain better representations for a subject's response. It is extremely helpful in single trial analysis cases~\cite{de2019multiway}. The assumption is that the MCCA models identify the stimuli related components from each response and remove all the components unrelated and subject-specific to the stimulus. 

\subsection{Impact of Hyperparameters} \label{subsec:hyper}
\color{black}

In this section, we analyze the impact of the hyperparameters involved in the deep CCA/MCCA models on the correlation metric.  The parameters analyzed are:  model architecture, dropout percentage, regularization parameter in DMCCA, and the number of output dimensions in DMCCA. We use a learning rate of $1e-3$ and a batch size of $2048$ on experiments where these parameters are not mentioned. Further, the number of time-lags used in the stimulus input is also varied to understand its impact. 
% For both the linear and deep models, they are varied over same range of values and the values with the best performance are selected.
For initializing the models, we start with multiple random seeds and choose the one which gives the best correlation on the validation data before the model training.  {\color{black} 
% The deep models used in the DCCA and DMCCA are specified in Section V-B. 
Unless specified otherwise, the values of parameters $d_s$ and $\lambda$ are fixed to $60$ and $0.1$. The value of $d$ is fixed to $1$ and $10$ for all the DCCA and DMCCA models respectively.}
% This procedure allows to find a better a local maxima in the training. 
% This helps to avoid the chance of obtaining seeds with bad initialization.
% This helps us to make sure that the model's performance is not entirely left to chance.

\subsubsection{Dropout}
\begin{figure*}
  \centering
  \includegraphics[width=18cm]{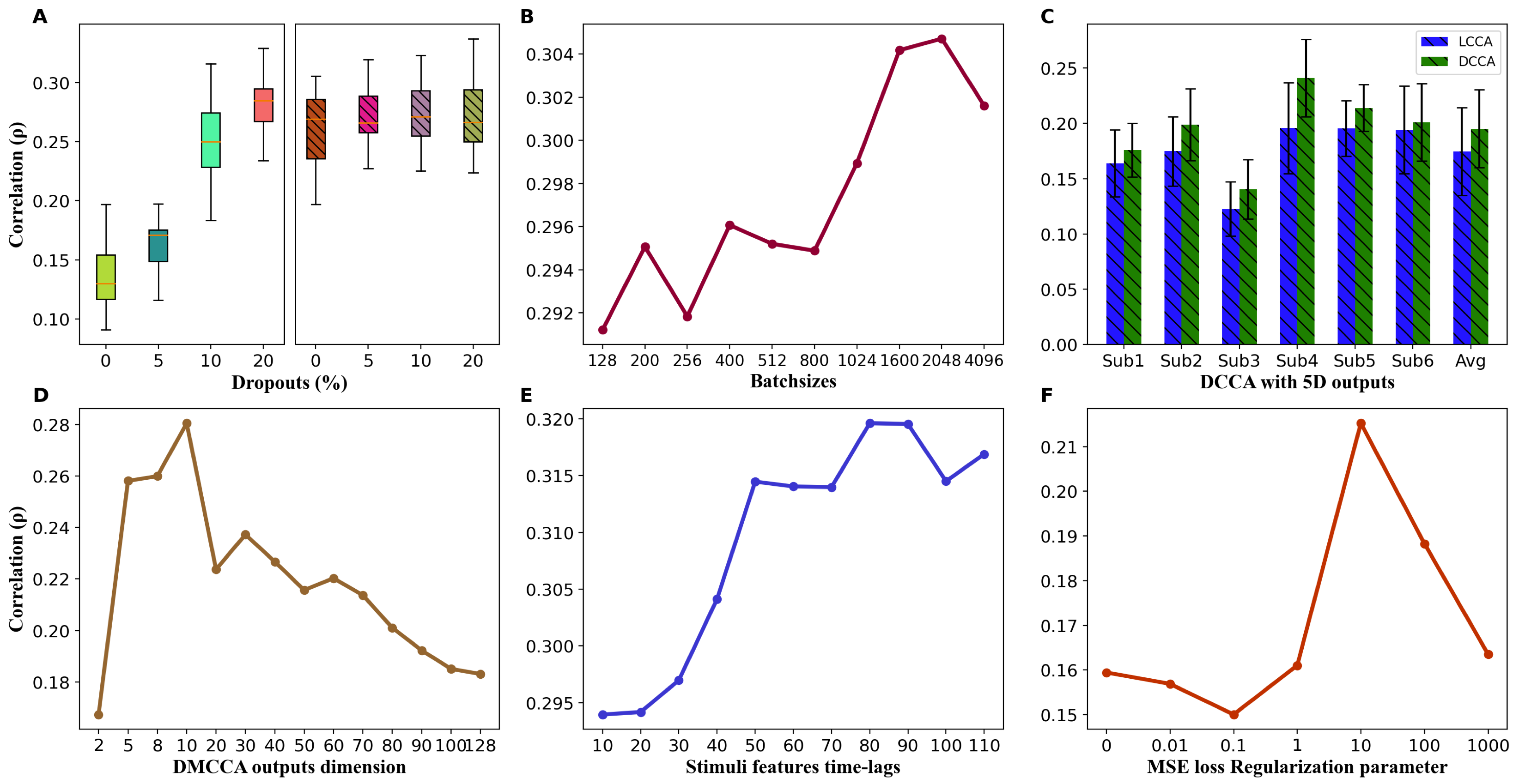}
  \caption{\textbf{A}. Effect of Dropout percentage on the final correlations of DMLC method and DCCA method. \textbf{B}. Effect of varying batchsizes on a DCCA model with output as $1$D. \textbf{C}. Comparison between the correlation per dimension of the final representations from linear and deep CCA models with outputs of $5$D. \textbf{D}. Comparing the performance of DMLC with varying output dimensionality  \textbf{E}. Impact of choice of time-lags for the stimulus, given as input to the DMLC method. \textbf{F}. Effect of the MSE regularization parameter in the DMLC method.}
    \label{fig:hypers}
\end{figure*}

% EEG data is very noisy i.e., SNR is too low. Hence, regularization techniques like dropout ~\cite{hinton2012improving,srivastava2014dropout} are incorporated in the deep models training to avoid over-fitting in these noisy conditions. 
For the speech-EEG dataset, we experiment with dropout percentage from $0-20$\% in the DCCA/DMCCA model. The correlation values obtained by DMLC and DCCA methods are shown in Figure~\ref{fig:hypers} (A). 
% The first four unshaded boxes correspond to the DMLC method's final correlation values. And the next four correspond to that of DCCA method. 
When there is no dropout, there is a tendency for the model to overfit. 
% Also, a higher value of dropout can result in an underfit as seen in this figure. 
A similar effect is seen in the DCCA model as well.

\subsubsection{Batchsize}
The effect of the batch size is analyzed for the DCCA model. The average correlation values of $6$ subjects from the speech dataset, for $20$ cross-validation trials is reported in Figure~\ref{fig:hypers} (B). Given the noisy nature of the data, we find that the higher batchsizes (compared to typical choices of few hundreds in supervised classification setting) are found to improve the final correlation value. The optimal batch size on the validation data is $2048$. 

\subsubsection{DCCA output dimension}
Just like the CCA model where multiple canonical components dimensions can be derived from the data, the DCCA model also can be trained for multiple output dimensions. The comparison of the linear and deep CCA models for $5$ canonical correlation components is shown in  Figure~\ref{fig:hypers} (C) for each subject in the speech-EEG dataset. As seen here, the DCCA model improves over the linear CCA model consistently for all the subjects.  

\subsubsection{DMCCA output dimension}
We have varied the number of output dimensions in the DMCCA model from $10$ to $128$. The performance, average of the final correlation values of all the subjects, of the DMLC model is presented in the Figure~\ref{fig:hypers} (D) on the speech-EEG dataset. The best performance is achieved for DMCCA outputs of $10$ dimensions.

\subsubsection{Context size of stimuli}
The time-lag applied to the stimulus ($d_s$) in the DMCCA model also plays an important role. It is varied from $10$ to $110$. For $6$ subjects from the speech-EEG dataset, the DMLC model is trained and tested and the average correlation of all the subjects is shown in the Figure~\ref{fig:hypers} (E).   A stimulus time-lag of $80$ gives the best performance.

\begin{figure}[t!]
  \centering
  \includegraphics[width=8.6cm]{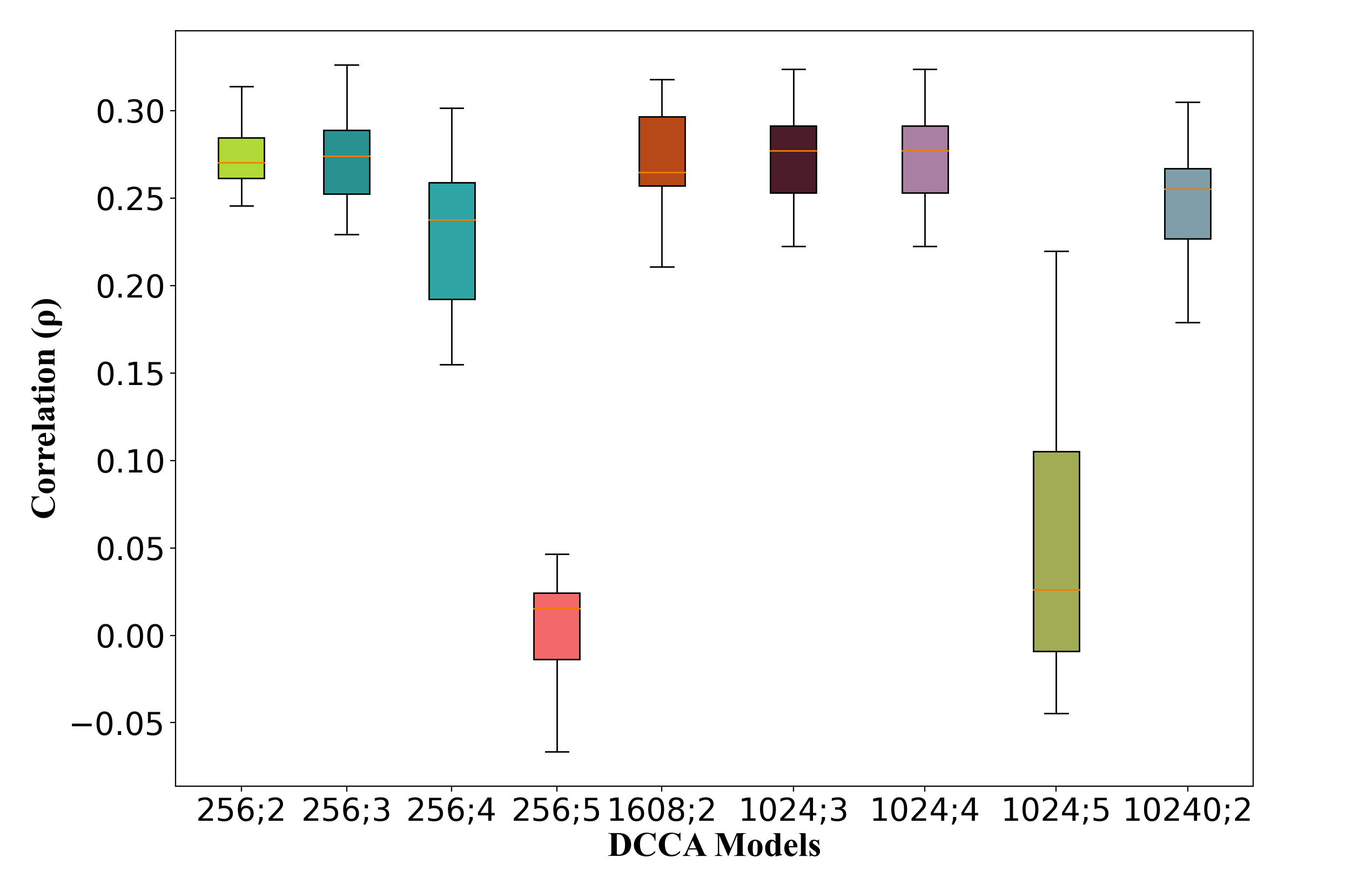}
  \caption{Different architectures explored in DCCA models. The x-axis denotes ``number of units per layer; number of layers"}.
   \label{fig:diff_archs}
   \vspace{-0.5cm}
\end{figure}

\subsubsection{Regularization parameter}
% Two deep variants of Linear MCCA model are tried: DGCCA and DMCCA. 
The DMCCA model incorporates the MSE regularization loss in addition to the inter-set correlation loss. The effect of the regularization parameter ($\lambda$) is studied in Figure~\ref{fig:hypers} (F), by varying it from from $0$ to $1000$. The results show that the regularization with $\lambda = 10$  gives the best performance. 
% All the results provided are from this proposed model, with the best regularization parameter. 
% The ISC of the outputs of a DMCCA model, with same initialization, varying the strength of the regularization is shown in figure. ~\ref{fig:hypers}. 
% Though the regularization term helps to provide better correlated denoised outputs, the final outputs of LCCA3 do not vary much.

% \begin{figure}[t!]
%   % \framebox{\includegraphics[width=8.6cm]{lmdcca3.png}}
%   \includegraphics[width=8.8cm]{diff_archs.png}
%   \centering
%   \caption{Different architectures tried in deep versions of CCA3 and MCCA models.}
%   \label{fig:diff_archs}
% \end{figure}

\subsection{Model Architecture}
We also experimented with various architecture choices for the DCCA model. The experiments with varying the number of hidden layers ($L$) from $2$ to $5$ and number of units ($n_L$) in each layer is shown in Figure~\ref{fig:diff_archs}. As seen in this plot, increasing the number of layers degrades the correlation, as the model tends to over-fit. {\color{black} This trend may also be attributed to the lack of sufficient audio-EEG data for each subject.  We hypothesize that, with more training data, the deeper models may further improve over linear models as well as the shallow models.}

\color{black}

\subsection{Impact of Improved Correlations}
The EEG recordings capture various unrelated brain processes along with the response to the stimuli. Thus, only a fraction of the variance in the EEG signal can be explained by its stimulus. This results in low correlation values for many of the linear methods proposed in the past.   In this paper, we have explored the application of deep models   whereby consistent improvements in correlations are illustrated.  Many applications based on EEG would benefit significantly with the improvements in stimulus-response correlations. For example, the improved correlations will help 
% in the development of an 
the EEG enabled auditory assistance device (e.g. hearing aid) as suggested by Cheveigné et al.~\cite{de2018decoding}.  %The improvement of decoding algorithms will also enable the understanding of attentional effects in listening to cocktail party speech.  
In music information retrieval, the performance improvement in EEG decoding systems will enable the understanding of the perceptual attributes of music. 
%For example, models can be built using the transformed outputs for classifying music genre as well as familiarity/likeness of musical snippets.  
Throughout the study, we have pursued simple features like envelope. However, audio signals are encoded in several other dimensions like pitch, rhythm, zero-crossings, phase, semantic/linguistic features etc. The exploration of the model with additional features may further throw light on the encoding of these dimensions in the EEG signals. Further, the techniques   proposed in this work would be applicable to other brain signals like MEG, ECoG and fMRI signals as well.

\section{Summary}
\label{sec:summary}
In this paper, we have proposed extensions to models that uncover the stimulus-response relationships between auditory signals like speech and music and their EEG responses. The models advance linear correlation methods and are proposed for improving single trial analyses. The models pose the problem of finding the optimal transforms that need to be applied to the stimulus and response in a deep learning framework which enables the learning of these transforms using established optimization methods. Using the proposed deep models, we show that the correlations between the stimulus and the response can be significantly improved over the linear methods. Further, the applicability of the proposed models is separately analyzed for speech and music EEG tasks.  

\section*{Acknowledgements}
This work started from initial discussions in the Telluride neuromorphic workshop 2019 with  Alain De Cheveigné.  We would  like to acknowledge the early efforts on deep correlation analysis performed by Sandeep Kothinti and Malcolm Slaney. 
We also thank Michael Broderick for the speech-EEG dataset and Blair Kaneshiro for the NMED-H dataset. 

\bibliographystyle{IEEEtran}
\bibliography{refer}

% Various experiment setup:

% Linear CCA for each subject separately. \\
% Deep CCA for each subject separately. \\

% To get help from other subjects' repsonse data. First naive trial would be: \\
% Linear CCA for all subjects at a time. \\
% Deep CCA for all subjects at a time. \\

% Already proved methods which successfully obtain help from other response data. \\
% Linear MCCA for all response data + LCCA3/DCCA3 for each subject separately. \\
% Deep MCCA for all response data + LCCA3/DCCA3 for each subject separately. 
%\newpage

\newpage

\appendix
\subsection{Derivation of the deep CCA model gradients}\label{sec:appendixA}

The matrix $\T _H $ is defined as $\T_H \triangleq \hat{\C}_{\x\x}^{-1/2}\hat{\C}_{\x\y}\hat{\C}_{\y\y}^{-1/2}$ and its SVD decomposition is denoted as $\T_H = \U\D\V^{\top}$. The objective function that needs to be maximized is 
\vspace{-16pt}
\begin{equation}
    \rho\left(\H_\x, \H_\y\right) = \|\T _H\|_\text{tr} =  \operatorname{Trace}\left(\T _H^{\top}\T_H\right)^{1/2} \nonumber
\vspace{-16pt}
\end{equation}
% $\rho\left(\H_\x, \H_\y\right) = \|\T _H\|_\text{tr} =  \operatorname{Trace}\left(\T _H^{\top}\T_H\right)^{1/2}$ \hfill \\
where $\|\cdot\|_\text{tr}$ denotes the trace norm of the matrix.

\textcolor{black}{
We can derive the partial derivative of the objective function with respect to the matrix $\hat{\C}_{\x\x}$, $\nabla_{\x\x}$, as following:}
{\color{black}
\begin{align}
\left[\nabla_{\x\x}\right]^{i j} &= \frac{\partial \rho\left(\H_\x, \H_\y\right)} {\partial\left[\hat{\C}_{\x\x}\right]^{i j}} \nonumber \\ 
&= \sum_{k l} \frac{\partial \rho\left(\H_\x, \H_y\right)} {\partial \left[\T_H^{\top}\T_H\right]^{k l}} \cdot \frac{\partial \left[\T_H^{\top}\T_H\right]^{k l} } {\partial\left[\hat{\C}_{\x\y}\right]^{i j}} \nonumber \\
&= \sum_{k l}\left[\frac{1}{2} \left(\T_H^{\top}\T_H\right)^{-1/2} \right]^{k l} \cdot \frac{\partial \left[\T_H^{\top}\T_H\right]^{k l} } {\partial\left[\hat{\C}_{\x\y}\right]^{i j}} \label{eq:last}
\end{align}
where,
\begin{align}
    &\frac{\partial \left[\T_H^{\top}\T_H\right]^{k l} } {\partial\left[\hat{\C}_{\x\y}\right]^{i j}} = \sum_{p q} \frac{\partial\left[\T_H^{\top} \T_H\right]^{k l} } {\partial\left[\hat{\C}_{\x\x}^{-1}\right]^{p q}} \frac{\partial\left[\hat{\C}_{\x\x}^{-1}\right]^{p q} }{\partial\left[\hat{\C}_{\x\x}\right]^{i j}} \nonumber \\
    &= \sum_{p q} \frac{ \partial \left[\hat{\C}_{\y\y}^{-1/2} \hat{\C}_{\y\x} \hat{\C}_{\x\x}^{-1} \hat{\C}_{\x\y} \hat{\C}_{\y\y}^{-1/2}\right]^{k l}} {\partial\left[\hat{\C}_{\x\x}^{-1} \right]^{p q}} \frac{\partial\left[\hat{\C}_{\x\x}^{-1}\right]^{p q} }{\partial\left[\hat{\C}_{\x\x}\right]^{i j}} \nonumber \\
    &= -\sum_{p q} \left[\hat{\C}_{\y\y}^{-1/2} \hat{\C}_{\y\x}\right]^{k p} \left[\hat{\C}_{\x\y} \hat{\C}_{\y\y}^{-1/2}\right]^{q l} \left[\hat{\C}_{\x\x}^{-1}\right]^{p i} \left[\hat{\C}_{\x\x}^{-1}\right]^{j q} \nonumber \\
    &= -\left[\hat{\C}_{\y\y}^{-1/2} \hat{\C}_{\y\x} \hat{\C}_{\x\x}^{-1}\right]^{k i} \left[\hat{\C}_{\x\x}^{-1} \hat{\C}_{\x\y} \hat{\C}_{\y\y}^{-1/2}\right]^{j l} \nonumber \\
    &=-\left[\T_H^{\top} \hat{\C}_{\x\x}^{-1/2}\right]^{k i} \left[\hat{\C}_{\x\x}^{-1 / 2} \T_H\right]^{j l}
\end{align}
Now, Equation~(\ref{eq:last}) can be rewritten as :
\begin{align}
&\left[\nabla_{\x\x}\right]^{i j} \nonumber \\
&=-\frac{1}{2} \sum_{k l}\left[\T_H^{\top} \hat{\C}_{\x\x}^{-1/2} \right]^{k i} \left[ \left(\T_H^{\top}\T_H\right)^{-1/2} \right]^{k l} \left[\hat{\C}_{\x\x}^{-1 / 2} \T_H\right]^{j l}  \nonumber \\ 
&=-\frac{1}{2} \sum_{k l}\left[\hat{\C}_{\x\x}^{-1 / 2} \T_H\right]^{i k} \left[\left(\T_H^{\top} \T_H\right)^{-1 / 2}\right]^{k l} \left[\T_H^{\top} \hat{\C}_{\x\x}^{-1 / 2}\right]^{l j}  \nonumber \\
&=-\frac{1}{2}\left[\hat{\C}_{\x\x}^{-1 / 2} \T_H\left(\T_H^{\top} \T_H\right)^{-1 / 2} \T_H^{\top} \hat{\C}_{\x\x}^{-1 / 2}\right]^{i j}  \nonumber \\
&=-\frac{1}{2} \left[\hat{\C}_{\x\x}^{-1 / 2} \U \D \V^{\top}\left(\V \D^{-1} \V^{\top}\right) \V \D \U^{\top} \hat{\C}_{\x\x}^{-1 / 2}\right]^{i j}  \nonumber \\
&=-\frac{1}{2} \left[\hat{\C}_{\x\x}^{-1 / 2} \U \D \U^{\top} \hat{\C}_{\x\x}^{-1 / 2}\right]^{i j} 
\end{align}

Similarly, the partial derivative of the objective function with respect to the matrix $\hat{\C}_{\x\y}$, $\nabla_{\x\y}$, can be derived as:
\begin{align}
\left[\nabla_{\x\y}\right]^{i j} &= \frac{\partial \rho\left(\H_\x, \H_\y\right)} {\partial\left[\hat{\C}_{\x\y}\right]^{i j}}  \nonumber \\
&= \sum_{k l} \frac{\partial \rho\left(\H_\x, \H_y\right)} {\partial \left[\T_H\right]^{k l}} \cdot \frac{\partial \left[\T_H\right]^{k l}} {\partial\left[\hat{\C}_{\x\y}\right]^{i j}}  \nonumber \\
&= \sum_{k l}\left[\U \V^{\top}\right]^{k l} \cdot \left[\hat{\C}_{\x\x}^{-1 / 2}\right]^{k i} \left[\hat{\C}_{\y\y}^{-1 / 2}\right]^{j l}  \nonumber \\
&= \left[\hat{\C}_{\x\x}^{-1 / 2} \U \V^{\top} \hat{\C}_{\y\y}^{-1 / 2}\right]^{i j}
\end{align}

% Hence, the two equations~(\ref{eq:delxy}) and~(\ref{eq:delxx}) can be derived using the above procedure. 

Now, the Equation~(\ref{grads}) can be derived as following. Let us denote the gradient of the objective function $\rho\left(\H_\x, \H_\y\right)$ with respect to $\H_\x$ as :
\begin{equation}
    \frac{\partial \rho\left(\H_\x, \H_\y\right)}{\partial \left[ \H_\x \right]^{i j}} =\sum_{k l} \left[\nabla_{\x\x}\right]^{k l} \frac{\partial \left[ \hat{\C}_{\x\x}\right]^{k l} } {\partial \left[ \H_\x\right]^{i j}} + \sum_{k l} \left[\nabla_{\x\y}\right]^{k l} \frac{\partial \left[ \hat{\C}_{\x\y}\right]^{k l} } {\partial \left[ \H_\x\right]^{i j}}  \label{eq:delHH}
\end{equation}
Each term from the right hand side can be derived as:
\begin{align}
    &\frac{\partial \left[\hat{\C}_{\x\x}\right]^{ab}}{ \partial \left[\H_{\x}\right]^{i j}} = \frac{1}{m-1} \frac{\partial } { \partial \left[\H_{\x}\right]^{i j}} \left( \left(\left[\bar{\H}_\x\right]^{a*}\right)^{\top} \left[\bar{\H}_\x\right]^{b*} \right) \nonumber  \\
    % = \frac{1}{m-1}\frac{\partial \sum_{k}{\left[\bar{\H_\x}\right]^{a k}}{\left[\bar{\H_\x}\right]^{k b}}}{ \partial \left[\H_{\x}\right]^{i j}} \nonumber  \\
    &= \left\{\begin{array}{ll} \frac{2}{m-1}\left(\left[\H_\x\right]^{i j}-\frac{1}{m} \sum_{k} \left[\H_\x\right]^{i k}\right) & \text { if } a=i, b=i \nonumber  \\ 
    \frac{1}{m-1}\left(\left[\H_\x\right]^{b j}-\frac{1}{m} \sum_{k} \left[\H_\x\right]^{b k}\right) & \text { if } a=i, b \neq i  \nonumber \\ 
    \frac{1}{m-1}\left(\left[\H_\x\right]^{a j}-\frac{1}{m} \sum_{k} \left[\H_\x\right]^{a k}\right) & \text { if } a \neq i, b=i  \nonumber \\ 
    0 & \text { if } a \neq i, b \neq i\end{array}\right.  \nonumber \\
    &= \frac{1}{m-1}\left(1_{\{a=i\}} \left[\bar{\H}_\x\right]^{b j} + 1_{\{b=i\}} \left[\bar{\H}_\x\right]^{a j}\right)
\end{align}\\
And, 
\begin{align} 
\frac{\partial \left[\hat{\C}_{\x\y}\right]^{a b}}{\partial \left[\H_\x\right]^{i j}} &=\frac{1}{m-1} 1_{\{a=i\}}\left(\left[\H_\y\right]^{b j} - \frac{1}{m} \sum_{k} \left[\H_\y\right]^{b k}\right)  \nonumber \\ 
&=\frac{1}{m-1} 1_{\{a=i\}} \left[\bar{\H}_\y\right]^{b j} 
\end{align}\\
Hence, equation~(\ref{eq:delHH}) can be rewritten as:
\begin{align}
    &\frac{\partial \rho\left(\H_\x, \H_\y\right)}{\partial \left[ \H_\x \right]^{i j}} \nonumber \\
    &= \sum_{k l} \left[\nabla_{\x\x}\right]^{k l} \frac{\partial \left[ \hat{\C}_{\x\x}\right]^{k l} } {\partial \left[ \H_\x\right]^{i j}} + \sum_{k l} \left[\nabla_{\x\y}\right]^{k l} \frac{\partial \left[ \hat{\C}_{\x\y}\right]^{k l} } {\partial \left[ \H_\x\right]^{i j}}  \nonumber \\
    &= \frac{1}{m-1}\left(\sum_{l} \nabla_{\x\x}^{i l} \bar{\H}_\x^{l j} + \sum_{k} \nabla_{\x\x}^{k i} \bar{\H}_\x^{k j} + \sum_{l} \nabla_{\x\y}^{i l} \bar{H}_\y^{l j}\right)  \nonumber \\ 
    &=\frac{1}{m-1}\left(\left[\nabla_{\x\x} \bar{\H}_{\x}\right]^{i j} + \left[\nabla_{\x\x}^{\top} \bar{\H}_{\x}\right]^{i j} + \left[\nabla_{\x\y} \bar{\H}_{\y}\right]^{i j}\right)  \nonumber \\
    &=\frac{1}{m-1}\left( 2 \left[\nabla_{\x\x} \bar{\H}_{\x}\right]^{i j} + \left[\nabla_{\x\y} \bar{\H}_{\y}\right]^{i j}\right)
\end{align}

Therefore, 
\begin{equation}
    \frac{\partial \rho\left(\mathbf{H_x}, \mathbf{H_y}\right)}{\partial \mathbf{H_x}}=\frac{1}{m-1}\left(2 \mathbf{\nabla_{\mathbf{x}\mathbf{x}}} \mathbf{\bar{H}_x}+\mathbf{\nabla_{\mathbf{x}\mathbf{y}}} \mathbf{\bar{H}_y}\right)
\end{equation}

}

\subsection{Acoustic features used for NMED-H Dataset}

{\color{black}
The $20$ acoustic features are extracted from the NMED-H Dataset as discussed in the baseline work by Gang et al.~\cite{gang2017decoding}. The features are extracted using the MIR toolbox provided by Lartillot et al.~\cite{lartillot2007matlab}. The $20$ acoustic features are calculated as following. \\

Let, $M_t\left[f\right]$ represent the magnitude of Discrete Fourier transform  (DFT) of a given audio signal ($m_t\left(n\right)$) at frame instant $t$ and frequency bin $f$. Let, the number of frequency bins be $F$. 
\begin{enumerate}
    \item \textbf{Zero Crossing Rate}: It represents the number of sign changes of the audio signal $m_t(n)$.
    \item \textbf{Spectral centroid}: The spectral centroid is the first order moment of the DFT  given as :
    \begin{equation}
        C_t = \frac{\sum_{f=1}^F fM_t\left[f\right]}{\sum_{f=1}^F M_t\left[f\right]}
    \end{equation}
    \item \textbf{High/Low Energy Ratio}: It represents the ratio of the highest to the lowest magnitudes in $M_t\left[f\right]$.
    \item \textbf{Spectral Spread}: It represents the standard deviation of the $M_t\left[f\right]$ in the frequency domain.
    \item \textbf{Spectral Roll-off}:  At a given instant, the roll-off $R_t$ is measured as the frequency below which $85$\% of the magnitude of the Fourier transform is concentrated.
    \begin{equation}
        \sum _{f=1}^{R_t} M_t[f] = 0.85 \sum_{f=1}^F M_t\left[f\right]
    \end{equation}
    \item \textbf{Spectral Entropy}: It is measured as the relative Shannon entropy of the normalized magnitude $M_t\left[f\right]$.
    \item \textbf{Spectral Flatness}: It represents whether the magnitude distribution in the frequency domain is smooth or spiky. It is measured as the ratio between the geometric mean and the arithmetic mean of the magnitudes in the frequency domain at each instant. 
    \begin{equation}
        F_t = \frac{\sqrt[F]{\pi_{f=1}^{F} M_t\left[f\right] }}{\frac{1}{F}\sum_{f=1}^F M_t\left[f\right]}
    \end{equation}
    \item \textbf{Roughness}: It tries to measure the sensory distance related to the beating phenomenon when a pair of sinusoidal signals are close in frequency. It is estimated by obtaining the peaks of the DFT  $M_t\left[f\right]$, and taking the average of  the distances between all  pairs of peaks.
    \item \textbf{RMS energy}: It measures the root mean square value of the magnitudes in the frequency domain, $M_t\left[f\right]$, at a given instant.
    \item \textbf{Broadband Spectral Flux}: It measures the Euclidean distance between the normalized DFT and its previous instant $M_{t-1}[f]$.
    \item \textbf{Spectral flux for $10$ octave-wide sub-bands}: It is measured by dividing the frequency domain into $10$ bins and calculating the absolute differences of each bin for the two time instants (after normalizing the DFT). 
    \begin{equation}
        S_t[f] = \left(M_{t-1}\left[f\right] - M_t\left[f\right]\right)^2
    \end{equation}
    
\end{enumerate}

More details about these features are available in the primer of the music information retrieval (MIR) toolbox by Lartillot et al.~\cite{lartillot2007matlab}.

}

\end{document}